\documentclass{article}%
\usepackage{amssymb}
\usepackage{amsmath}
\usepackage{graphicx}
\usepackage{amsfonts}%
\providecommand{\U}[1]{\protect\rule{.1in}{.1in}}
\begin{document}

\title{On classical finite probability theory as a quantum probability calculus}
\author{David Ellerman\\Department of Philosophy \\U. of California/Riverside}
\maketitle

\begin{abstract}
This paper shows how the classical finite probability theory (with
equiprobable outcomes) can be reinterpreted and recast as the quantum
probability calculus of a pedagogical or "toy" model of \textit{quantum
mechanics over sets} (QM/sets). There have been four previous attempts to
develop a quantum-like model with the base field of $%
%TCIMACRO{\U{2102} }%
%BeginExpansion
\mathbb{C}
%EndExpansion
$ replaced by $%
%TCIMACRO{\U{2124} }%
%BeginExpansion
\mathbb{Z}
%EndExpansion
_{2}$, but they are all forced into a merely "modal" interpretation by
requiring the brackets to take values in $%
%TCIMACRO{\U{2124} }%
%BeginExpansion
\mathbb{Z}
%EndExpansion
_{2}$ ($1=$ possible, $0=$ impossible). But the usual QM brackets
$\left\langle \psi|\varphi\right\rangle $ give the "overlap" between states
$\psi$ and $\varphi$, so for sets $S,T\subseteq U$, the natural definition is
$\left\langle S|T\right\rangle =\left\vert S\cap T\right\vert $. This allows
QM/sets to be developed with a full probability calculus that turns out to be
the perfectly classical Laplace-Boole finite probability theory. The point is
not to clarify finite probability theory but to elucidate quantum mechanics
itself by seeing some of its quantum features (e.g., two-slit experiment) in a
classical setting.

\end{abstract}
\tableofcontents

\section{Introduction}

This paper develops a pedagogical or "toy" model of quantum mechanics over
sets (QM/sets) where the quantum probability calculus is the ordinary
Laplace-Boole finite logical probability theory (\cite{laplace:probs},
\cite{boole:lot}) and where the usual vector spaces over $%
%TCIMACRO{\U{2102} }%
%BeginExpansion
\mathbb{C}
%EndExpansion
$ for QM are replaced with vector spaces over $%
%TCIMACRO{\U{2124} }%
%BeginExpansion
\mathbb{Z}
%EndExpansion
_{2}$ in QM/sets. Quantum mechanics over sets is a bare-bones "logical" (e.g.,
non-physical\footnote{In full QM, the DeBroglie relations connect mathematical
notions such as frequency and wave-length to physical notions such as energy
and momentum. QM/sets is "non-physical" in the sense that it is a sets-version
of the pure mathematical framework of (finite-dimensional) QM without those
direct physical connections.}) version of QM with appropriate versions of
spectral decomposition, the Dirac brackets, ket-bra resolution, the norm,
observable-attributes, and the Born rule all in the simple classical setting
of sets, but that nevertheless provides models of characteristically quantum
results (e.g., a QM/sets version of the double-slit experiment). In that
manner, QM/sets can serve not only as a pedagogical (or "toy") model of QM but
perhaps as an engine to better elucidate QM itself by representing the quantum
features in a simple classical setting.

There have been at least four previous attempts at developing a version of QM
over sets, i.e., where the base field of $%
%TCIMACRO{\U{2102} }%
%BeginExpansion
\mathbb{C}
%EndExpansion
$ is replaced by $%
%TCIMACRO{\U{2124} }%
%BeginExpansion
\mathbb{Z}
%EndExpansion
_{2}$ (\cite{schum:modal}, \cite{hansonsabry:dqt}, \cite{tak:mutant}, and
\cite{abram-coecke:catqm}). All these attempts use the aspect of full QM that
the brackets and the observables take their values in the base field. When the
base field is $%
%TCIMACRO{\U{2124} }%
%BeginExpansion
\mathbb{Z}
%EndExpansion
_{2}$, then the models do "not make use of the idea of probability"\cite[p.
919]{schum:modal} and have instead only a modal interpretation ($1=$
possibility and $0=$ impossibility).

The model of QM over sets developed here is based on a different understanding
of the relation between the pedagogical model and full QM. Instead of trying
to mimic QM (replacing $%
%TCIMACRO{\U{2102} }%
%BeginExpansion
\mathbb{C}
%EndExpansion
$ with $%
%TCIMACRO{\U{2124} }%
%BeginExpansion
\mathbb{Z}
%EndExpansion
_{2}$), the idea is that QM/sets can perfectly well have the brackets and
observables take values \textit{outside} the base field of $%
%TCIMACRO{\U{2124} }%
%BeginExpansion
\mathbb{Z}
%EndExpansion
_{2}$ (e.g., use real-valued observables = real-valued random variables in
classical finite probability theory) and even defining a more primitive
version of "eigenvectors" and "eigenvalues" that are not (in general) the
eigenvectors and eigenvalues of linear operators on the vector space over $%
%TCIMACRO{\U{2124} }%
%BeginExpansion
\mathbb{Z}
%EndExpansion
_{2}$. The transitioning from QM/sets to full QM is then seen not as going
from one model to another model of a set of axioms (e.g., as in
\cite{abram-coecke:catqm}) but as a process of "internalization" allowed by
increasing the base field from $%
%TCIMACRO{\U{2124} }%
%BeginExpansion
\mathbb{Z}
%EndExpansion
_{2}$ to $%
%TCIMACRO{\U{2102} }%
%BeginExpansion
\mathbb{C}
%EndExpansion
$. The increased power of $%
%TCIMACRO{\U{2102} }%
%BeginExpansion
\mathbb{C}
%EndExpansion
$ (e.g., algebraic completeness) then allows the primitive "eigenvectors" and
"eigenvalues" of QM/sets to be "internalized" as true eigenvectors and
eigenvalues of (Hermitian) linear operators on vector spaces over $%
%TCIMACRO{\U{2102} }%
%BeginExpansion
\mathbb{C}
%EndExpansion
$ and the brackets can then also be "internalized" as a bilinear inner product
taking values in the base field $%
%TCIMACRO{\U{2102} }%
%BeginExpansion
\mathbb{C}
%EndExpansion
$. Hence under this approach (and in contrast to the four previous
approaches), the "taking values in the base field" is seen \textit{only} as an
aspect of full QM over $%
%TCIMACRO{\U{2102} }%
%BeginExpansion
\mathbb{C}
%EndExpansion
$ and not as a necessary aspect of a pedagogical proto-QM model such as
QM/sets with the base field of $%
%TCIMACRO{\U{2124} }%
%BeginExpansion
\mathbb{Z}
%EndExpansion
_{2}$.

\section{Laplace-Boole probability theory}

Since our purpose is conceptual rather than mathematical, we will stick to the
simplest case of finite probability theory with a finite sample space
$U=\left\{  u_{1},...,u_{n}\right\}  $ of $n$ equiprobable outcomes and to
finite dimensional QM.\footnote{The mathematics can be generalized to the case
where each point $u_{i}$ in the sample space has a probability $p_{i}$ but the
simpler case of equiprobable points serves our conceptual purposes.} The
\textit{events} are the subsets $S\subseteq U$, and the \textit{probability}
of an event $S$ occurring in a trial is the ratio of the cardinalities:
$\Pr\left(  S\right)  =\frac{\left\vert S\right\vert }{\left\vert U\right\vert
}$. Given that a conditioning event $S\subseteq U$ occurs, the
\textit{conditional probability} that $T\subseteq U$ occurs is: $\Pr
(T|S)=\frac{\Pr\left(  T\cap S\right)  }{\Pr(S)}=\frac{\left\vert T\cap
S\right\vert }{\left\vert S\right\vert }$. The ordinary probability
$\Pr\left(  T\right)  $ of an event $T$ can be taken as the conditional
probability with $U$ as the conditioning event so all probabilities can be
seen as conditional probabilities. Given a (real-valued) random variable,
i.e., a \textit{numerical attribute} $f:U\rightarrow%
%TCIMACRO{\U{211d} }%
%BeginExpansion
\mathbb{R}
%EndExpansion
$ on the elements of \ $U$, the \textit{probability of observing a value }%
$r$\textit{ given an event }$S$\textit{ }is the conditional probability of the
event $f^{-1}\left(  r\right)  $ given $S$:

\begin{center}
$\Pr\left(  r|S\right)  =\frac{\left\vert f^{-1}\left(  r\right)  \cap
S\right\vert }{\left\vert S\right\vert }$.
\end{center}

\noindent That is all the probability theory we will need here. Our task is to
show how the mathematics of finite probability theory can be recast using the
mathematical notions of quantum mechanics with the base field of $%
%TCIMACRO{\U{2124} }%
%BeginExpansion
\mathbb{Z}
%EndExpansion
_{2}$ is substituted, \textit{mutatis mutandis}, for the complex numbers $%
%TCIMACRO{\U{2102} }%
%BeginExpansion
\mathbb{C}
%EndExpansion
$.

\section{Recasting finite probability theory as a quantum probability
calculus}

\subsection{Vector spaces over $%
%TCIMACRO{\U{2124} }%
%BeginExpansion
\mathbb{Z}
%EndExpansion
_{2}$}

To show how classical Laplace-Boole finite probability theory can be recast as
a quantum probability calculus, we use finite dimensional vector spaces over $%
%TCIMACRO{\U{2124} }%
%BeginExpansion
\mathbb{Z}
%EndExpansion
_{2}$.\footnote{We are assuming some basic familarity with the mathematics of
finite dimensional QM.} The power set $\wp\left(  U\right)  $ of $U=\left\{
u_{1},...,u_{n}\right\}  $ is a vector space over $%
%TCIMACRO{\U{2124} }%
%BeginExpansion
\mathbb{Z}
%EndExpansion
_{2}=\left\{  0,1\right\}  $, isomorphic to $%
%TCIMACRO{\U{2124} }%
%BeginExpansion
\mathbb{Z}
%EndExpansion
_{2}^{n}$, where the vector addition $S+T$ is the \textit{symmetric
difference} (or inequivalence) of subsets. That is, for $S,T\subseteq U$,

\begin{center}
$S+T=\left(  S-T\right)  \cup\left(  T-S\right)  =S\cup T-S\cap T$.
\end{center}

\noindent The $U$\textit{-basis} in $\wp\left(  U\right)  $ is the set of
singletons $\left\{  u_{1}\right\}  ,\left\{  u_{2}\right\}  ,...,\left\{
u_{n}\right\}  $, i.e., the set $\left\{  \left\{  u\right\}  \right\}  _{u\in
U}$. A vector $S\in\wp\left(  U\right)  $ is specified in the $U$-basis as
$S=\sum_{u\in S}\left\{  u\right\}  $ and it is characterized by its $%
%TCIMACRO{\U{2124} }%
%BeginExpansion
\mathbb{Z}
%EndExpansion
_{2}$-valued characteristic function $\chi_{S}:U\rightarrow%
%TCIMACRO{\U{2124} }%
%BeginExpansion
\mathbb{Z}
%EndExpansion
_{2}\subseteq%
%TCIMACRO{\U{211d} }%
%BeginExpansion
\mathbb{R}
%EndExpansion
$ of coefficients since $S=\sum_{u\in U}\chi_{S}\left(  u\right)  \left\{
u\right\}  $. Similarly, a vector $v$ in $%
%TCIMACRO{\U{2102} }%
%BeginExpansion
\mathbb{C}
%EndExpansion
^{n}$ is specified in terms of an orthonormal basis $\left\{  \left\vert
v_{i}\right\rangle \right\}  $ as $v=\sum_{i}c_{i}\left\vert v_{i}%
\right\rangle $ and is characterized by a $%
%TCIMACRO{\U{2102} }%
%BeginExpansion
\mathbb{C}
%EndExpansion
$-valued function $\left\langle \_|v\right\rangle :\left\{  v_{i}\right\}
\rightarrow%
%TCIMACRO{\U{2102} }%
%BeginExpansion
\mathbb{C}
%EndExpansion
$ assigning a complex amplitude $\left\langle v_{i}|v\right\rangle =c_{i}$ to
each basis vector $\left\vert v_{i}\right\rangle $.

One of the key pieces of mathematical machinery in QM, namely the inner
product, does not exist in vector spaces over finite fields but brackets can
still be defined starting with $\left\langle \left\{  u\right\}
|_{U}S\right\rangle =\chi_{S}\left(  u\right)  $ (see below) and a norm can be
defined to play a similar role in the probability calculus of QM/sets.

Seeing $\wp\left(  U\right)  $ as the abstract vector space $%
%TCIMACRO{\U{2124} }%
%BeginExpansion
\mathbb{Z}
%EndExpansion
_{2}^{n}$ allows different bases in which the vectors can be expressed (as
well as the basis-free notion of a vector as a "ket"). Hence the quantum
probability calculus developed here can be seen as a "non-commutative"
generalization of the classical Laplace-Boole finite probability theory where
a different basis corresponds to a different equicardinal sample space
$U^{\prime}=\left\{  u_{1}^{\prime},...,u_{n}^{\prime}\right\}  $.

Consider the simple case of $U=\left\{  a,b,c\right\}  $ where the $U$-basis
is $\left\{  a\right\}  $, $\left\{  b\right\}  $, and $\left\{  c\right\}  $.
The three subsets $\left\{  a,b\right\}  $, $\left\{  b,c\right\}  $, and
$\left\{  a,b,c\right\}  $ also form a basis since:

$\left\{  b,c\right\}  +\left\{  a,b,c\right\}  =\left\{  a\right\}  $;

$\left\{  b,c\right\}  +\left\{  a,b\right\}  +\left\{  a,b,c\right\}
=\left\{  b\right\}  $; and

$\left\{  a,b\right\}  +\left\{  a,b,c\right\}  =\left\{  c\right\}  $.

\noindent These new basis vectors could be considered as the basis-singletons
in another equicardinal universe $U^{\prime}=\left\{  a^{\prime},b^{\prime
},c^{\prime}\right\}  $ where $\left\{  a^{\prime}\right\}  $, $\left\{
b^{\prime}\right\}  $, and $\left\{  c^{\prime}\right\}  $ refer to the same
abstract vector as $\left\{  a,b\right\}  $, $\left\{  b,c\right\}  $, and
$\left\{  a,b,c\right\}  $ respectively.

In the following \textit{ket table}, each row is an abstract vector of $%
%TCIMACRO{\U{2124} }%
%BeginExpansion
\mathbb{Z}
%EndExpansion
_{2}^{3}$ expressed in the $U$-basis, the $U^{\prime}$-basis, and a
$U^{\prime\prime}$-basis.

\begin{center}%
\begin{tabular}
[c]{|c|c|c|}\hline
$U=\left\{  a,b,c\right\}  $ & $U^{\prime}=\left\{  a^{\prime},b^{\prime
},c^{\prime}\right\}  $ & $U^{\prime\prime}=\left\{  a^{\prime\prime
},b^{\prime\prime},c^{\prime\prime}\right\}  $\\\hline\hline
$\left\{  a,b,c\right\}  $ & $\left\{  c^{\prime}\right\}  $ & $\left\{
a^{\prime\prime},b^{\prime\prime},c^{\prime\prime}\right\}  $\\\hline
$\left\{  a,b\right\}  $ & $\left\{  a^{\prime}\right\}  $ & $\left\{
b^{\prime\prime}\right\}  $\\\hline
$\left\{  b,c\right\}  $ & $\left\{  b^{\prime}\right\}  $ & $\left\{
b^{\prime\prime},c^{\prime\prime}\right\}  $\\\hline
$\left\{  a,c\right\}  $ & $\left\{  a^{\prime},b^{\prime}\right\}  $ &
$\left\{  c^{\prime\prime}\right\}  $\\\hline
$\left\{  a\right\}  $ & $\left\{  b^{\prime},c^{\prime}\right\}  $ &
$\left\{  a^{\prime\prime}\right\}  $\\\hline
$\left\{  b\right\}  $ & $\left\{  a^{\prime},b^{\prime},c^{\prime}\right\}  $
& $\left\{  a^{\prime\prime},b^{\prime\prime}\right\}  $\\\hline
$\left\{  c\right\}  $ & $\left\{  a^{\prime},c^{\prime}\right\}  $ &
$\left\{  a^{\prime\prime},c^{\prime\prime}\right\}  $\\\hline
$\emptyset$ & $\emptyset$ & $\emptyset$\\\hline
\end{tabular}

Vector space isomorphism: $%
%TCIMACRO{\U{2124} }%
%BeginExpansion
\mathbb{Z}
%EndExpansion
_{2}^{3}\cong\wp\left(  U\right)  \cong\wp\left(  U^{\prime}\right)  \cong%
\wp\left(  U^{\prime\prime}\right)  $ where row = ket.
\end{center}

In the Dirac notation \cite{dirac:principles}, the\textit{ ket} $\left\vert
\left\{  a,c\right\}  \right\rangle $ represents the abstract vector that is
represented in the $U$-basis as $\left\{  a,c\right\}  $. A row of the ket
table gives the different representations of the \textit{same} ket in the
different bases, e.g., $\left\vert \left\{  a,c\right\}  \right\rangle
=\left\vert \left\{  a^{\prime},b^{\prime}\right\}  \right\rangle =\left\vert
\left\{  c^{\prime\prime}\right\}  \right\rangle $.

\subsection{The brackets}

In a Hilbert space, the inner product is used to define the brackets
$\left\langle v_{i}|v\right\rangle $ and the norm $\left\vert v\right\vert
=\sqrt{\left\langle v|v\right\rangle }$. In a vector space over $%
%TCIMACRO{\U{2124} }%
%BeginExpansion
\mathbb{Z}
%EndExpansion
_{2}$, the Dirac notation can still be used to define the brackets and norm
even though there is no inner product. For a singleton basis vector $\left\{
u\right\}  \subseteq U$, the \textit{bra} $\left\langle \left\{  u\right\}
\right\vert _{U}:\wp\left(  U\right)  \rightarrow%
%TCIMACRO{\U{211d} }%
%BeginExpansion
\mathbb{R}
%EndExpansion
$ is defined by the \textit{bracket}:

\begin{center}
$\left\langle \left\{  u\right\}  |_{U}S\right\rangle =\left\{
\begin{array}
[c]{c}%
1\text{ if }u\in S\\
0\text{ if }u\notin S
\end{array}
\right.  =\left\vert \left\{  u\right\}  \cap S\right\vert =\chi_{S}\left(
u\right)  $.
\end{center}

\noindent Note that the bra and the bracket is defined in terms of the
$U$-basis and that is indicated by the $U$-subscript on the bra portion of the
bracket. Then for $u_{i},u_{j}\in U$, $\left\langle \left\{  u_{i}\right\}
|_{U}\left\{  u_{j}\right\}  \right\rangle =\chi_{\left\{  u_{j}\right\}
}\left(  u_{i}\right)  =\chi_{\left\{  u_{i}\right\}  }\left(  u_{j}\right)
=\delta_{ij}$ (the Kronecker delta function) which is the QM/sets-version of
$\left\langle v_{i}|v_{j}\right\rangle =\delta_{ij}$ for an orthonormal basis
$\left\{  \left\vert v_{i}\right\rangle \right\}  $ of $%
%TCIMACRO{\U{2102} }%
%BeginExpansion
\mathbb{C}
%EndExpansion
^{n}$. The bracket linearly extends in the reals to any two vectors $T,S\in
\wp\left(  U\right)  $:\footnote{\noindent Here $\left\langle T|_{U}%
S\right\rangle =\left\vert T\cap S\right\vert $ takes values outside the base
field of $%
%TCIMACRO{\U{2124} }%
%BeginExpansion
\mathbb{Z}
%EndExpansion
_{2}$ just like, say, the Hamming distance function $d_{H}\left(  T,S\right)
=\left\vert T+S\right\vert $ on vector spaces over $%
%TCIMACRO{\U{2124} }%
%BeginExpansion
\mathbb{Z}
%EndExpansion
_{2}$ in coding theory. \cite{mceliece:coding} Thus the bra $\left\langle
S\right\vert _{U}$ is not to be confused with the dual functional $\chi
_{S}:\wp\left(  U\right)  \rightarrow%
%TCIMACRO{\U{2124} }%
%BeginExpansion
\mathbb{Z}
%EndExpansion
_{2}$ that does take values in the base field. The brackets taking values in
the base field is a consequence of the base field being strengthened to $%
%TCIMACRO{\U{2102} }%
%BeginExpansion
\mathbb{C}
%EndExpansion
$. It is not a necessary feature of a quantum probability calculus as we see
in QM/sets.}

\begin{center}
$\left\langle T|_{U}S\right\rangle =\left\vert T\cap S\right\vert $.
\end{center}

\noindent This is the QM/sets-version of the Dirac brackets in the mathematics
of QM.

For more motivation, consider an orthonormal basis set $\left\{  \left\vert
v_{i}\right\rangle \right\}  _{i=1,...,n}$ in an $n$-dimensional Hilbert space
$V$ and the association $\left\{  u_{i}\right\}  \leftrightarrow$ $\left\vert
v_{i}\right\rangle $ for $i=1,...,n$. Given two subsets $T,S\subseteq U$,
$T=\sum_{u_{i}\in T}\left\{  u_{i}\right\}  $ corresponds to the unnormalized
$\psi_{T}=\sum_{u_{i}\in T}\left\vert v_{i}\right\rangle $ and similarly for
$\psi_{S}$. Then their inner product (defined using the $\left\{  \left\vert
v_{i}\right\rangle \right\}  _{i=1,...,n}$ basis) in $V$ is $\left\langle
\psi_{T}|\psi_{S}\right\rangle =\left\vert T\cap S\right\vert =\left\langle
T|_{U}S\right\rangle $. In both cases, the bracket gives a measure of the
overlap or indistinctness of the two vectors.\footnote{Indeed in QM/sets, the
brackets $\left\langle T|_{U}S\right\rangle =\left\vert T\cap S\right\vert $
for $T,T^{\prime},S\subseteq U$ should be thought of \textit{only} as a
measure of the overlap since they are not even linear, e.g., $\left\langle
T+T^{\prime}|_{U}S\right\rangle \neq\left\langle T|_{U}S\right\rangle
+\left\langle T^{\prime}|_{U}S\right\rangle $ whenever $T\cap T^{\prime}%
\neq\emptyset$. Only as the base field $%
%TCIMACRO{\U{2124} }%
%BeginExpansion
\mathbb{Z}
%EndExpansion
_{2}$ is increased to $%
%TCIMACRO{\U{2102} }%
%BeginExpansion
\mathbb{C}
%EndExpansion
$ (or $%
%TCIMACRO{\U{211d} }%
%BeginExpansion
\mathbb{R}
%EndExpansion
$) do the brackets 'fall into place' as a bilinear inner product. QM/sets is
not 'supposed' to have completely the same mathematical structure as QM only
with $%
%TCIMACRO{\U{2124} }%
%BeginExpansion
\mathbb{Z}
%EndExpansion
_{2}$ replacing $%
%TCIMACRO{\U{2102} }%
%BeginExpansion
\mathbb{C}
%EndExpansion
$. QM/sets is a proto-QM where things only 'fall into place' and are
'internalized' as the transition is made from $%
%TCIMACRO{\U{2124} }%
%BeginExpansion
\mathbb{Z}
%EndExpansion
_{2}$ to $%
%TCIMACRO{\U{2102} }%
%BeginExpansion
\mathbb{C}
%EndExpansion
$ as the base field.}

\subsection{Ket-bra resolution}

The \textit{ket-bra} $\left\vert \left\{  u\right\}  \right\rangle
\left\langle \left\{  u\right\}  \right\vert _{U}$ is defined as the
one-dimensional projection operator:

\begin{center}
$\left\vert \left\{  u\right\}  \right\rangle \left\langle \left\{  u\right\}
\right\vert _{U}=\left\{  u\right\}  \cap():\wp\left(  U\right)
\rightarrow\wp\left(  U\right)  $
\end{center}

\noindent and the \textit{ket-bra identity} holds as usual:

\begin{center}
$\sum_{u\in U}\left\vert \left\{  u\right\}  \right\rangle \left\langle
\left\{  u\right\}  \right\vert _{U}=\sum_{u\in U}\left(  \left\{  u\right\}
\cap()\right)  =I:\wp\left(  U\right)  \rightarrow\wp\left(  U\right)  $
\end{center}

\noindent where the summation is the symmetric difference of sets in
$\wp\left(  U\right)  $ and $I$ is the identity map [as a linear operator on
$\wp\left(  U\right)  $]. The overlap $\left\langle T|_{U}S\right\rangle $ can
be resolved using the ket-bra identity in the same basis:

\begin{center}
$\left\langle T|_{U}S\right\rangle =\sum_{u}\left\langle T|_{U}\left\{
u\right\}  \right\rangle \left\langle \left\{  u\right\}  |_{U}S\right\rangle
$.
\end{center}

\noindent Similarly a ket $\left\vert S\right\rangle $ for $S\subseteq U$ can
be resolved in the $U$-basis;

\begin{center}
$\left\vert S\right\rangle =\sum_{u\in U}\left\vert \left\{  u\right\}
\right\rangle \left\langle \left\{  u\right\}  |_{U}S\right\rangle =\sum_{u\in
U}\left\langle \left\{  u\right\}  |_{U}S\right\rangle \left\vert \left\{
u\right\}  \right\rangle =\sum_{u\in U}\left\vert \left\{  u\right\}  \cap
S\right\vert \left\vert \left\{  u\right\}  \right\rangle $
\end{center}

\noindent where a subset $S\subseteq U$ is just expressed as the sum of the
singletons $\left\{  u\right\}  \subseteq S$. That is \textit{ket-bra
resolution} in QM/sets. The ket $\left\vert S\right\rangle $ is the same as
the ket $\left\vert S^{\prime}\right\rangle $ for some subset $S^{\prime
}\subseteq U^{\prime}$ in another $U^{\prime}$-basis, but when the bra
$\left\langle \left\{  u\right\}  \right\vert _{U}$ is applied to the ket
$\left\vert S\right\rangle =\left\vert S^{\prime}\right\rangle $, then it is
the subset $S\subseteq U$, not $S^{\prime}\subseteq U^{\prime}$, that comes
outside the ket symbol $\left\vert {}\right\rangle $\noindent\ in
$\left\langle \left\{  u\right\}  |_{U}S\right\rangle =\left\vert \left\{
u\right\}  \cap S\right\vert $.\footnote{The term "$\left\{  u\right\}  \cap
S^{\prime}$" is not even defined since it is the intersection of subsets
$\left\{  u\right\}  \subseteq U$ and $S^{\prime}\subseteq U^{\prime}$ of two
different universe sets $U$ and $U^{\prime}$.}

\subsection{The norm}

The $U$\textit{-norm} $\left\Vert S\right\Vert _{U}:\wp\left(  U\right)
\rightarrow%
%TCIMACRO{\U{211d} }%
%BeginExpansion
\mathbb{R}
%EndExpansion
$ is defined, as usual, as the square root of the bracket:\footnote{We use the
double-line notation $\left\Vert S\right\Vert _{U}$ for the $U$-norm of a set
to distinguish it from the single-line notation $\left\vert S\right\vert $ for
the cardinality of a set, whereas the customary absolute value notation for
the norm of a vector $v$ in ordinary QM is $\left\vert v\right\vert
=\sqrt{\left\langle v|v\right\rangle }$. The context should suffice to
distinguish $\left\vert S\right\vert $ from $\left\vert v\right\vert $.}

\begin{center}
$\left\Vert S\right\Vert _{U}=\sqrt{\left\langle S|_{U}S\right\rangle }%
=\sqrt{\left\vert S\cap S\right\vert }=\sqrt{|S|}$
\end{center}

\noindent for $S\in\wp\left(  U\right)  $ which is the QM/sets-version of the
norm $\left\vert \psi\right\vert =\sqrt{\left\langle \psi|\psi\right\rangle }$
in ordinary QM. Note that a ket has to be expressed in the $U$-basis to apply
the $U$-norm definition so, for example, $\left\Vert \left\{  a^{\prime
}\right\}  \right\Vert _{U}=\sqrt{2}$ since $\left\vert \left\{  a^{\prime
}\right\}  \right\rangle =\left\vert \left\{  a,b\right\}  \right\rangle $.

\subsection{Numerical attributes and linear operators}

In classical physics, the observables are numerical attributes, e.g., the
assignment of a position and momentum to particles in phase space. One of the
differences between classical and quantum physics is the replacement of these
observable numerical attributes by linear operators associated with the
observables where the values of the observables appear as eigenvalues of the
operators. But this difference may be smaller than it would seem at first
since a numerical attribute $f:U\rightarrow%
%TCIMACRO{\U{211d} }%
%BeginExpansion
\mathbb{R}
%EndExpansion
$ can be recast into an operator-like format in QM/sets, and there is even a
QM/sets-analogue of spectral decomposition.

An observable, i.e., a Hermitian operator, on a Hilbert space $V$ has a home
basis set of orthonormal eigenvectors. In a similar manner, a real-valued
attribute $f:U\rightarrow%
%TCIMACRO{\U{211d} }%
%BeginExpansion
\mathbb{R}
%EndExpansion
$ defined on $U$ has the $U$-basis as its "home basis set." The connection
between the numerical attributes $f:U\rightarrow%
%TCIMACRO{\U{211d} }%
%BeginExpansion
\mathbb{R}
%EndExpansion
$ of QM/sets and the Hermitian operators of full QM can be established by
seeing the function $f$ as being \textit{like} an "operator" $f\upharpoonright
()$ on $\wp\left(  U\right)  $ in that it is used to define a sets-version of
an "eigenvalue" equation [where $f\upharpoonright S$ is the
\textit{restriction} of $f$ to $S\in\wp\left(  U\right)  $]. For any subset
$S\in\wp\left(  U\right)  $, the definition of the equation is:

\begin{center}
$f\upharpoonright S=rS$ holds iff $f$ is constant on the subset $S$ with the
value $r$.
\end{center}

\noindent This is the QM/sets-version of an \textit{eigenvalue equation} for
numerical attributes $f:U\rightarrow%
%TCIMACRO{\U{211d} }%
%BeginExpansion
\mathbb{R}
%EndExpansion
$. Whenever $S$ satisfies $f\upharpoonright S=rS$ for some $r$, then $S$ is
said to be an \textit{eigenvector} in the vector space $\wp\left(  U\right)  $
of the numerical attribute $f:U\rightarrow%
%TCIMACRO{\U{211d} }%
%BeginExpansion
\mathbb{R}
%EndExpansion
$, and $r\in%
%TCIMACRO{\U{211d} }%
%BeginExpansion
\mathbb{R}
%EndExpansion
$ is the associated \textit{eigenvalue}. Each eigenvalue $r$ determines as
usual an \textit{eigenspace} $\wp\left(  f^{-1}\left(  r\right)  \right)  $ of
its eigenvectors which is a subspace of the vector space $\wp\left(  U\right)
$. The whole space $\wp\left(  U\right)  $ can be expressed as usual as the
direct sum of the eigenspaces: $\wp\left(  U\right)  =%
%TCIMACRO{\tbigoplus \nolimits_{r\in f\left(  U\right)  }}%
%BeginExpansion
{\textstyle\bigoplus\nolimits_{r\in f\left(  U\right)  }}
%EndExpansion
\wp\left(  f^{-1}\left(  r\right)  \right)  $. Moreover, for distinct
eigenvalues $r\neq r^{\prime}$, any corresponding eigenvectors $S\in\wp\left(
f^{-1}\left(  r\right)  \right)  $ and $T\in\wp\left(  f^{-1}\left(
r^{\prime}\right)  \right)  $ are \textit{orthogonal} in the sense that
$\left\langle T|_{U}S\right\rangle =0$. In general, for vectors $S,T\in
\wp\left(  U\right)  $, orthogonality means zero overlap, i.e., disjointness.

The characteristic function $\chi_{S}:U\rightarrow%
%TCIMACRO{\U{211d} }%
%BeginExpansion
\mathbb{R}
%EndExpansion
$ for $S\subseteq U$ has the eigenvalues of $0$ and $1$ so it is a numerical
attribute that \textit{can} be "internalized" as a linear operator
$S\cap():\wp\left(  U\right)  \rightarrow\wp\left(  U\right)  $. Hence in this
case, the "eigenvalue equation" $f\upharpoonright T=rT$ for $f=\chi_{S}$
becomes an actual eigenvalue equation $S\cap T=rT$ for a linear\footnote{It
should be noted that the projection operator $S\cap():\wp\left(  U\right)
\rightarrow\wp\left(  U\right)  $ is not only idempotent but linear, i.e.,
$\left(  S\cap T_{1}\right)  +(S\cap T_{2})=S\cap\left(  T_{1}+T_{2}\right)
$. Indeed, this is the distributive law when $\wp\left(  U\right)  $ is
interpreted as a Boolean ring with intersection as multiplication.} operator
$S\cap()$ with the resulting eigenvalues of $1$ and $0$, and with the
resulting eigenspaces $\wp\left(  S\right)  $ and $\wp\left(  S^{c}\right)  $
(where $S^{c}$ is the complement of $S$) agreeing with those "eigenvalues" and
"eigenspaces" defined above for an arbitrary numerical attribute
$f:U\rightarrow%
%TCIMACRO{\U{211d} }%
%BeginExpansion
\mathbb{R}
%EndExpansion
$. The characteristic attributes $\chi_{S}:U\rightarrow%
%TCIMACRO{\U{211d} }%
%BeginExpansion
\mathbb{R}
%EndExpansion
$ are characterized by the property that their value-wise product, i.e.,
$\left(  \chi_{S}\bullet\chi_{S}\right)  \left(  u\right)  =\chi_{S}\left(
u\right)  \chi_{S}\left(  u\right)  $, is equal to the attribute value
$\chi_{S}\left(  u\right)  $, and that is reflected in the idempotency of the
corresponding operators:

\begin{center}
$\wp\left(  U\right)  \overset{S\cap()}{\longrightarrow}\wp\left(  U\right)
\overset{S\cap()}{\longrightarrow}\wp\left(  U\right)  =\wp\left(  U\right)
\overset{S\cap()}{\longrightarrow}\wp\left(  U\right)  $.
\end{center}

\noindent Thus the operators $S\cap()$ corresponding to the characteristic
attributes $\chi_{S}$ are \textit{projection operators}.\footnote{In order for
general real-valued attributes to be internalized as linear operators, in the
way that characteristic functions $\chi_{S}$ were internalized as projection
operators $S\cap()$, the base field would have to be strengthened to $%
%TCIMACRO{\U{2102} }%
%BeginExpansion
\mathbb{C}
%EndExpansion
$ and that would take us, \textit{mutatis mutandis}, from the probability
calculus of QM/sets to that of full QM.}

The (maximal) eigenvectors $f^{-1}\left(  r\right)  $ for $f$, with $r$ in the
\textit{image} or \textit{spectrum} $f\left(  U\right)  \subseteq%
%TCIMACRO{\U{211d} }%
%BeginExpansion
\mathbb{R}
%EndExpansion
$, span the set $U$, i.e., $U=\sum_{r\in f\left(  U\right)  }f^{-1}\left(
r\right)  $. Hence the attribute $f:U\rightarrow%
%TCIMACRO{\U{211d} }%
%BeginExpansion
\mathbb{R}
%EndExpansion
$ has a spectral decomposition in terms of its (projection-defining)
characteristic functions:

\begin{center}
$f=\sum_{r\in f\left(  U\right)  }r\chi_{f^{-1}\left(  r\right)
}:U\rightarrow%
%TCIMACRO{\U{211d} }%
%BeginExpansion
\mathbb{R}
%EndExpansion
$

\textit{Spectral decomposition} of set attribute $f:U\rightarrow%
%TCIMACRO{\U{211d} }%
%BeginExpansion
\mathbb{R}
%EndExpansion
$
\end{center}

\noindent which is the QM/sets-version of the spectral decomposition
$L=\sum_{\lambda}\lambda P_{\lambda}$ of a Hermitian operator $L$ in terms of
the projection operators $P_{\lambda}$ for its eigenvalues $\lambda$.

\subsection{Completeness and orthogonality of projection operators}

For any vector $S\in\wp\left(  U\right)  $, the operator $S\cap():\wp\left(
U\right)  \rightarrow\wp\left(  U\right)  $ is the linear projection operator
to the subspace $\wp\left(  S\right)  \subseteq\wp\left(  U\right)  $. The
usual completeness and orthogonality conditions on projection operators
$P_{\lambda}$ to the eigenspaces of an observable-operator have
QM/sets-versions for numerical attributes $f:U\rightarrow%
%TCIMACRO{\U{211d} }%
%BeginExpansion
\mathbb{R}
%EndExpansion
$:

\begin{enumerate}
\item completeness: $\sum_{\lambda}P_{\lambda}=I:V\rightarrow V$ in QM has the QM/sets-version:
\end{enumerate}

\begin{center}
$\sum_{r}f^{-1}\left(  r\right)  \cap()=I:\wp\left(  U\right)  \rightarrow
\wp\left(  U\right)  $, and
\end{center}

\begin{enumerate}
\item[2.] orthogonality: for $\lambda\neq\mu$, $V\overset{P_{\mu
}}{\longrightarrow}V\overset{P_{\lambda}}{\longrightarrow}%
V=V\overset{0}{\longrightarrow}V$ (where $0$ is the zero operator) has the
QM/sets-version: for $r\neq r^{\prime}$,
\end{enumerate}

\begin{center}
$\wp\left(  U\right)  \overset{f^{-1}\left(  r^{\prime}\right)  \cap
()}{\longrightarrow}\wp\left(  U\right)  \overset{f^{-1}\left(  r\right)
\cap()}{\longrightarrow}\wp\left(  U\right)  =\wp\left(  U\right)
\overset{0}{\longrightarrow}\wp\left(  U\right)  $.
\end{center}

Note that in spite of the lack of an inner product, the orthogonality of
projection operators $S\cap()$ is perfectly well-defined in QM/sets where it
boils down to the disjointness of subsets, i.e., the cardinality of subsets'
overlap (instead of their inner product) being $0$.

\subsection{The Born Rule for measurement in QM and QM/sets}

An orthogonal decomposition of a finite set $U$ is just a partition
$\pi=\left\{  B\right\}  $ of $U$ since the blocks $B,B^{\prime},...$ are
orthogonal (i.e., disjoint) and their sum is $U$. Given such an orthogonal
decomposition of $U$, we have the:

\begin{center}
$\left\Vert U\right\Vert _{U}^{2}=\sum_{B\in\pi}\left\Vert B\right\Vert
_{U}^{2}$

Pythagorean Theorem

for orthogonal decompositions of sets.
\end{center}

An old question is: "why the squaring of amplitudes in the Born rule of QM?" A
superposition state between certain definite orthogonal alternatives $A$ and
$B$, where the latter are represented by vectors $\overrightarrow{A}$ and
$\overrightarrow{B}$, is represented by the vector sum $\overrightarrow{C}%
=\overrightarrow{A}+\overrightarrow{B}$. But what is the "strength,"
"intensity," or relative importance of the vectors $\overrightarrow{A}$ and
$\overrightarrow{B}$ in the vector sum $\overrightarrow{C}$? That question
requires a \textit{scalar} measure of strength or intensity. The magnitude or
"length" given by the norm $\left\Vert {}\right\Vert $ does not answer the
question since $\left\Vert \overrightarrow{A}\right\Vert +\left\Vert
\overrightarrow{B}\right\Vert \neq\left\Vert \overrightarrow{C}\right\Vert $.
But the Pythagorean Theorem shows that the norm-squared gives the scalar
measure of "intensity" that answers the question: $\left\Vert
\overrightarrow{A}\right\Vert ^{2}+\left\Vert \overrightarrow{B}\right\Vert
^{2}=\left\Vert \overrightarrow{C}\right\Vert ^{2}$ in vector spaces over $%
%TCIMACRO{\U{2124} }%
%BeginExpansion
\mathbb{Z}
%EndExpansion
_{2}$ or over $%
%TCIMACRO{\U{2102} }%
%BeginExpansion
\mathbb{C}
%EndExpansion
$. And when the superposition state is reduced by a measurement, then
the\textit{ probability} that the indefinite state will reduce to one of the
definite alternatives is given by that relative scalar measure of the
eigen-alternative's "strength" or "intensity" in the indefinite state--and
that is the Born Rule. In a slogan, Born is the off-spring of Pythagoras.

Given an orthogonal basis $\left\{  \left\vert v_{i}\right\rangle \right\}  $
in a finite dimensional Hilbert space and given the $U$-basis for the vector
space $\wp\left(  U\right)  $, the corresponding Pythagorean results for the
basis sets are:

\begin{center}
$\left\vert \psi\right\vert ^{2}=\sum_{i}\left\langle v_{i}|\psi\right\rangle
^{\ast}\left\langle v_{i}|\psi\right\rangle =\sum_{i}\left\vert \left\langle
v_{i}|\psi\right\rangle \right\vert ^{2}$ and

$\left\Vert S\right\Vert _{U}^{2}=\sum_{u\in U}\left\langle \left\{
u\right\}  |_{U}S\right\rangle ^{2}$.
\end{center}

Given an observable-operator in QM and a numerical attribute in QM/sets, the
corresponding Pythagorean Theorems for the complete sets of orthogonal
projection operators are:

\begin{center}
$\left\vert \psi\right\vert ^{2}=\sum_{\lambda}\left\vert P_{\lambda}\left(
\psi\right)  \right\vert ^{2}$ and

$\left\Vert S\right\Vert _{U}^{2}=\sum_{r}\left\Vert f^{-1}\left(  r\right)
\cap S\right\Vert _{U}^{2}=\sum_{r}\left\vert f^{-1}\left(  r\right)  \cap
S\right\vert =\left\vert S\right\vert $.
\end{center}

\noindent Normalizing gives:

\begin{center}
$\sum_{\lambda}\frac{\left\vert P_{\lambda}\left(  \psi\right)  \right\vert
^{2}}{\left\vert \psi\right\vert ^{2}}=1$ and

$\sum_{r}\frac{\left\Vert f^{-1}\left(  r\right)  \cap S\right\Vert _{U}^{2}%
}{\left\Vert S\right\Vert _{U}^{2}}=\sum_{r}\frac{\left\vert f^{-1}\left(
r\right)  \cap S\right\vert }{\left\vert S\right\vert }=1$
\end{center}

\noindent so the non-negative summands can be interpreted as
probabilities--which is the Born rule in QM and in QM/sets.\footnote{Note that
there is no notion of a normalized vector in a vector space over $%
%TCIMACRO{\U{2124} }%
%BeginExpansion
\mathbb{Z}
%EndExpansion
_{2}$ (another consequence of the lack of an inner product). The normalization
is, as it were, postponed to the probability algorithm which is computed in
the reals. This "external" probability algorithm is "internalized" when $%
%TCIMACRO{\U{2124} }%
%BeginExpansion
\mathbb{Z}
%EndExpansion
_{2}$ is strengthened to $%
%TCIMACRO{\U{2102} }%
%BeginExpansion
\mathbb{C}
%EndExpansion
$ in going from QM/sets to full QM.}

Here $\frac{\left\vert P_{\lambda}\left(  \psi\right)  \right\vert ^{2}%
}{\left\vert \psi\right\vert ^{2}}$ is the "mysterious" quantum probability of
getting $\lambda$ in an $L$-measurement of $\psi$, while $\frac{\left\Vert
f^{-1}\left(  r\right)  \cap S\right\Vert _{U}^{2}}{\left\Vert S\right\Vert
_{U}^{2}}=\frac{\left\vert f^{-1}\left(  r\right)  \cap S\right\vert
}{\left\vert S\right\vert }$ has the rather unmysterious interpretation in the
pedagogical model, QM/sets, as the probability $\Pr\left(  r|S\right)  $ of
the numerical attribute $f:U\rightarrow%
%TCIMACRO{\U{211d} }%
%BeginExpansion
\mathbb{R}
%EndExpansion
$ having the eigenvalue $r$ when "measuring" $S\in\wp\left(  U\right)  $. Thus
the QM/sets-version of the Born Rule is the perfectly ordinary Laplace-Boole
rule for the conditional probability $\Pr\left(  r|S\right)  =\frac{\left\vert
f^{-1}\left(  r\right)  \cap S\right\vert }{\left\vert S\right\vert }$, that
given $S\subseteq U$, a random variable $f:U\rightarrow%
%TCIMACRO{\U{211d} }%
%BeginExpansion
\mathbb{R}
%EndExpansion
$ takes the value $r$.

In QM/sets, when the indefinite state $S$ is being "measured" using the
observable $f$ where the probability $\Pr\left(  r|S\right)  $ of getting the
eigenvalue $r$ is $\frac{\left\Vert f^{-1}\left(  r\right)  \cap S\right\Vert
_{U}^{2}}{\left\Vert S\right\Vert _{U}^{2}}=\frac{\left\vert f^{-1}\left(
r\right)  \cap S\right\vert }{\left\vert S\right\vert }$, the "damned quantum
jump" (Schr\"{o}dinger) goes from $S$ by the projection operator
$f^{-1}\left(  r\right)  \cap()$ to the projected resultant state
$f^{-1}\left(  r\right)  \cap S$ which is in the eigenspace $\wp\left(
f^{-1}\left(  r\right)  \right)  $ for that eigenvalue $r$. The state
resulting from the measurement represents a more-definite state $f^{-1}\left(
r\right)  \cap S$ that now has the definite $f$-value of $r$--so a second
measurement would yield the same eigenvalue $r$ with probability:

\begin{center}
$\Pr\left(  r|f^{-1}\left(  r\right)  \cap S\right)  =\frac{\left\vert
f^{-1}\left(  r\right)  \cap\left[  f^{-1}\left(  r\right)  \cap S\right]
\right\vert }{\left\vert f^{-1}\left(  r\right)  \cap S\right\vert }%
=\frac{\left\vert f^{-1}\left(  r\right)  \cap S\right\vert }{\left\vert
f^{-1}\left(  r\right)  \cap S\right\vert }=1$
\end{center}

\noindent and the same resulting vector $f^{-1}\left(  r\right)  \cap\left[
f^{-1}\left(  r\right)  \cap S\right]  =f^{-1}\left(  r\right)  \cap S$ using
the idempotency of the projection operators.

Hence the treatment of measurement in QM/sets is \textit{all} analogous to the
treatment of measurement in standard Dirac-von-Neumann QM.

\subsection{Summary of QM/sets and QM}

The QM/set-versions of the corresponding QM notions are summarized in the
following table for the finite $U$-basis of the $%
%TCIMACRO{\U{2124} }%
%BeginExpansion
\mathbb{Z}
%EndExpansion
_{2}$-vector space $\wp\left(  U\right)  $ and for an orthonormal basis
$\left\{  \left\vert v_{i}\right\rangle \right\}  $ of a finite dimensional
Hilbert space $V$.

\begin{center}%
\begin{tabular}
[c]{|c|c|}\hline
QM/sets over $%
%TCIMACRO{\U{2124} }%
%BeginExpansion
\mathbb{Z}
%EndExpansion
_{2}$ & Standard QM over $%
%TCIMACRO{\U{2102} }%
%BeginExpansion
\mathbb{C}
%EndExpansion
$\\\hline\hline
Projections: $S\cap():\wp\left(  U\right)  \rightarrow\wp\left(  U\right)  $ &
$P:V\rightarrow V$ where $P^{2}=P$\\\hline
Spectral Decomposition.: $f=\sum_{r}r\chi_{f^{-1}\left(  r\right)  }$ &
$L=\sum_{\lambda}\lambda P_{\lambda}$\\\hline
Completeness.: $\sum_{r}f^{-1}\left(  r\right)  \cap()=I$ & $\sum_{\lambda
}P_{\lambda}=I$\\\hline
Orthog.: $r\neq r^{\prime}$, $\left[  f^{-1}\left(  r\right)  \cap()\right]
\left[  f^{-1}\left(  r^{\prime}\right)  \cap()\right]  =\emptyset\cap()$ &
$\lambda\neq\mu$, $P_{\lambda}P_{\mu}=0$\\\hline
Brackets: $\left\langle S|_{U}T\right\rangle =\left\vert S\cap T\right\vert $
= overlap of $S,T\subseteq U$ & $\left\langle \psi|\varphi\right\rangle =$
overlap of $\psi$ and $\varphi$\\\hline
Ket-bra: $\sum_{u\in U}\left\vert \left\{  u\right\}  \right\rangle
\left\langle \left\{  u\right\}  \right\vert _{U}=\sum_{u\in U}\left(
\left\{  u\right\}  \cap()\right)  =I$ & $\sum_{i}\left\vert v_{i}%
\right\rangle \left\langle v_{i}\right\vert =I$\\\hline
Resolution: $\left\langle S|_{U}T\right\rangle =\sum_{u}\left\langle
S|_{U}\left\{  u\right\}  \right\rangle \left\langle \left\{  u\right\}
|_{U}T\right\rangle $ & $\left\langle \psi|\varphi\right\rangle =\sum
_{i}\left\langle \psi|v_{i}\right\rangle \left\langle v_{i}|\varphi
\right\rangle $\\\hline
Norm: $\left\Vert S\right\Vert _{U}=\sqrt{\left\langle S|_{U}S\right\rangle
}=\sqrt{\left\vert S\right\vert }$ where $S\subseteq U$ & $\left\vert
\psi\right\vert =\sqrt{\left\langle \psi|\psi\right\rangle }$\\\hline
Basis Pythagoras: $\left\Vert S\right\Vert _{U}^{2}=\sum_{u\in U}\left\langle
\left\{  u\right\}  |_{U}S\right\rangle ^{2}=\left\vert S\right\vert $ &
$\left\vert \psi\right\vert ^{2}=\sum_{i}\left\langle v_{i}|\psi\right\rangle
^{\ast}\left\langle v_{i}|\psi\right\rangle $\\\hline
Normalized: $\sum_{u\in U}\frac{\left\langle \left\{  u\right\}
|_{U}S\right\rangle ^{2}}{\left\Vert S\right\Vert _{U}^{2}}=\sum_{u\in S}%
\frac{1}{\left\vert S\right\vert }=1$ & $\sum_{i}\frac{\left\langle v_{i}%
|\psi\right\rangle ^{\ast}\left\langle v_{i}|\psi\right\rangle }{\left\vert
\psi\right\vert ^{2}}=\sum_{i}\frac{\left\vert \left\langle v_{i}%
|\psi\right\rangle \right\vert ^{2}}{\left\vert \psi\right\vert ^{2}}%
=1$\\\hline
Basis Born rule: $\Pr\left(  \left\{  u\right\}  |S\right)  =\frac
{\left\langle \left\{  u\right\}  |_{U}S\right\rangle ^{2}}{\left\Vert
S\right\Vert _{U}^{2}}$ & $\Pr\left(  v_{i}|\psi\right)  =\frac{\left\vert
\left\langle v_{i}|\psi\right\rangle \right\vert ^{2}}{\left\vert
\psi\right\vert ^{2}}$\\\hline
Attribute Pythagoras: $\left\Vert S\right\Vert _{U}^{2}=\sum_{r}\left\Vert
f^{-1}\left(  r\right)  \cap S\right\Vert _{U}^{2}$ & $\left\vert
\psi\right\vert ^{2}=\sum_{\lambda}\left\vert P_{\lambda}\left(  \psi\right)
\right\vert ^{2}$\\\hline
Normalized: $\sum_{r}\frac{\left\Vert f^{-1}\left(  r\right)  \cap
S\right\Vert _{U}^{2}}{\left\Vert S\right\Vert _{U}^{2}}=\sum_{r}%
\frac{\left\vert f^{-1}\left(  r\right)  \cap S\right\vert }{\left\vert
S\right\vert }=1$ & $\sum_{\lambda}\frac{\left\vert P_{\lambda}\left(
\psi\right)  \right\vert ^{2}}{\left\vert \psi\right\vert ^{2}}=1$\\\hline
Attribute Born rule: $\Pr(r|S)=\frac{\left\Vert f^{-1}\left(  r\right)  \cap
S\right\Vert _{U}^{2}}{\left\Vert S\right\Vert _{U}^{2}}=\frac{\left\vert
f^{-1}\left(  r\right)  \cap S\right\vert }{\left\vert S\right\vert }$ &
$\Pr\left(  \lambda|\psi\right)  =\frac{\left\vert P_{\lambda}\left(
\psi\right)  \right\vert ^{2}}{\left\vert \psi\right\vert ^{2}}$\\\hline
\end{tabular}

Probability calculus for QM/sets over $%
%TCIMACRO{\U{2124} }%
%BeginExpansion
\mathbb{Z}
%EndExpansion
_{2}$ and for standard QM over $%
%TCIMACRO{\U{2102} }%
%BeginExpansion
\mathbb{C}
%EndExpansion
$
\end{center}

\section{Measurement in QM/sets}

\subsection{Measurement, Partitions, and Distinctions}

In QM/sets, numerical attributes $f:U\rightarrow%
%TCIMACRO{\U{211d} }%
%BeginExpansion
\mathbb{R}
%EndExpansion
$ can be considered as random variables on a set of equiprobable states
$\left\{  u\right\}  \subseteq U$. The inverse images of attributes (or random
variables) define set partitions $\left\{  f^{-1}\right\}  =\left\{
f^{-1}\left(  r\right)  \right\}  _{r\in f\left(  U\right)  }$ on the set $U$.
Considered abstractly, the partitions on a set $U$ are partially ordered by
refinement where a partition $\pi=\left\{  B\right\}  $ \textit{refines} a
partition $\sigma=\left\{  C\right\}  $, written $\sigma\preceq\pi$, if for
any block $B\in\pi$, there is a block $C\in\sigma$ such that $B\subseteq C$.
The principal logical operation needed here is the \textit{partition join}
where the join $\pi\vee\sigma$ is the partition whose blocks are the non-empty
intersections $B\cap C$ for $B\in\pi$ and $C\in\sigma$.

Each partition $\pi$ can be represented as a binary relation
$\operatorname*{dit}\left(  \pi\right)  \subseteq U\times U$ on $U$ where the
ordered pairs $\left(  u,u^{\prime}\right)  $ in $\operatorname*{dit}\left(
\pi\right)  $ are the \textit{distinctions} or \textit{dits} of $\pi$ in the
sense that $u$ and $u^{\prime}$ are in distinct blocks of $\pi$. These
\textit{dit sets} $\operatorname*{dit}\left(  \pi\right)  $ as binary
relations might be called \textit{partition relations} which are also called
"apartness relations" in computer science. An ordered pair $\left(
u,u^{\prime}\right)  $ is an \textit{indistinction} or \textit{indit} of $\pi$
if $u$ and $u^{\prime}$ are in the same block of $\pi$. The set of indits,
$\operatorname*{indit}\left(  \pi\right)  $, as a binary relation is just the
equivalence relation associated with the partition $\pi$, the complement of
the dit set $\operatorname*{dit}\left(  \pi\right)  $ in $U\times U$.

In the category-theoretic duality between \textit{sub}-sets (which are the
subject matter of Boole's subset logic, the latter being usually mis-specified
as the special case of "propositional" logic) and \textit{quotient}-sets or
partitions (\cite{ell:partitions} or \cite{ell:intropartlogic}), the
\textit{elements} of a subset and the \textit{distinctions} of a partition are
corresponding concepts.\footnote{Boole has been included along with Laplace in
the name of classical finite probability theory since he developed it as the
normalized counting measure on the elements of the subsets of his logic.
Applying the same mathematical move to the dual logic of partitions results in
developing the notion of \textit{logical entropy} $h\left(  \pi\right)  $ of a
partition $\pi$ as the normalized counting measure on the dit set
$\operatorname*{dit}\left(  \pi\right)  $, i.e., $h\left(  \pi\right)
=\frac{\left\vert \operatorname*{dit}\left(  \pi\right)  \right\vert
}{\left\vert U\times U\right\vert }$. (\cite{ell:distinctions},
\cite{ell:logentropy})}

The partial ordering of subsets in the Boolean lattice $\wp\left(  U\right)  $
is the inclusion of elements, and the refinement partial ordering of
partitions in the partition lattice $%
%TCIMACRO{\tprod }%
%BeginExpansion
{\textstyle\prod}
%EndExpansion
(U)$ is just the inclusion of distinctions, i.e., $\sigma\preceq\pi$ iff
$\operatorname*{dit}\left(  \sigma\right)  \subseteq\operatorname*{dit}\left(
\pi\right)  $. The top of the Boolean lattice is the subset $U$ of all
possible elements and the top of the partition lattice is the \textit{discrete
partition} $\mathbf{1}=\left\{  \left\{  u\right\}  \right\}  _{u\in U}$ of
singletons which makes all possible distinctions: $\operatorname*{dit}\left(
\mathbf{1}\right)  =U\times U-\Delta$ (where $\Delta=\left\{  \left(
u,u\right)  :u\in U\right\}  $ is the diagonal). The bottom of the Boolean
lattice is the empty set $\emptyset$ of no elements and the bottom of the
lattice of partitions is the \textit{indiscrete partition }(or \textit{blob})
$\mathbf{0}=\left\{  U\right\}  $ which makes no distinctions.

The two lattices can be illustrated in the case of $U=\left\{  a,b,c\right\}
$.%

%TCIMACRO{\FRAME{dtbpFU}{3.8232in}{1.8829in}{0pt}{\Qcb{Figure 1: Subset and
%partition lattices}}{}{fig1-two-lattices.eps}%
%{\special{ language "Scientific Word";  type "GRAPHIC";
%maintain-aspect-ratio TRUE;  display "USEDEF";  valid_file "F";
%width 3.8232in;  height 1.8829in;  depth 0pt;  original-width 3.6613in;
%original-height 1.7899in;  cropleft "0";  croptop "1";  cropright "1";
%cropbottom "0";  filename 'fig1-two-lattices.eps';file-properties "XNPEU";}}
%}%
%BeginExpansion
\begin{center}
\includegraphics[
height=1.8829in,
width=3.8232in
]%
{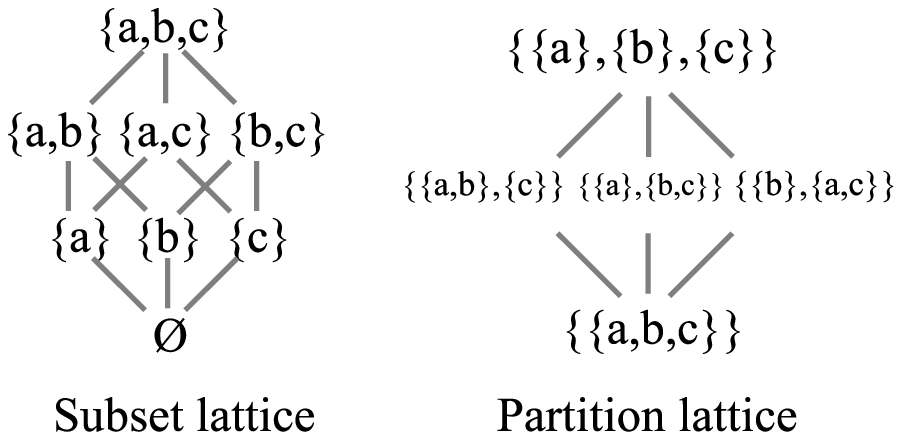}%
\\
Figure 1: Subset and partition lattices
\end{center}
%EndExpansion

In the correspondences between QM/sets and QM, a block $S$ in a partition on
$U$ [i.e., a vector $S\in\wp\left(  U\right)  $] corresponds to \textit{pure}
state in QM, and a partition $\pi=\left\{  B\right\}  $ on $U$ is the
\textit{mixed state} of orthogonal pure states $B$. In QM, a measurement makes
distinctions, i.e., makes alternatives distinguishable, and that turns a pure
state into a mixture of probabilistic outcomes. A measurement in QM/sets is
the distinction-creating process of turning a pure state $S\in\wp\left(
U\right)  $ into a mixed state partition $\left\{  f^{-1}\left(  r\right)
\cap S\right\}  _{r\in f\left(  U\right)  }$ on $S$. The distinction-creating
process of measurement in QM/sets is the action on $S$ of the inverse-image
partition $\left\{  f^{-1}\left(  r\right)  \right\}  _{r\in f\left(
U\right)  }$ in the join $\left\{  S,S^{c}\right\}  \vee\left\{  f^{-1}\left(
r\right)  \right\}  $ with the partition $\left\{  S,S^{c}\right\}  $, so that
action on $S$ is:

\begin{center}
$S\longrightarrow\left\{  f^{-1}\left(  r\right)  \cap S\right\}  _{r\in
f\left(  U\right)  }$

Action on the pure state $S$ of an $f$-measurement-join to give mixed state
$\left\{  f^{-1}\left(  r\right)  \cap S\right\}  _{r\in f\left(  U\right)  }$
on $S$.
\end{center}

\noindent The states $\left\{  f^{-1}\left(  r\right)  \cap S\right\}  _{r\in
f\left(  U\right)  }$ are all possible or "potential" but the actual
indefinite state $S$ turns into one of the definite states with the
probabilities given by the probability calculus: $\Pr(r|S)=\frac{\left\Vert
f^{-1}\left(  r\right)  \cap S\right\Vert _{U}^{2}}{\left\Vert S\right\Vert
_{U}^{2}}=\frac{\left\vert f^{-1}\left(  r\right)  \cap S\right\vert
}{\left\vert S\right\vert }$. Since the reduction of the state $S$ to the
state $f^{-1}\left(  r\right)  \cap S$ is mathematically described by applying
the projection operator $f^{-1}\left(  r\right)  \cap()$, it is called a
\textit{projective }measurement.

Hermann Weyl was at least one quantum theorist who touched on the relation
between what was, in effect, QM/sets and QM. He called a partition a "grating"
or "sieve," and then considered \textit{both} set partitions and vector space
partitions (direct sum decompositions) as the respective types of
gratings.\cite[pp. 255-257]{weyl:phil} He started with a numerical attribute
on a set, e.g., $f:U\rightarrow%
%TCIMACRO{\U{211d} }%
%BeginExpansion
\mathbb{R}
%EndExpansion
$, which defined the set partition or "grating" \cite[p. 255]{weyl:phil} with
blocks having the same attribute-value, e.g., $\left\{  f^{-1}\left(
r\right)  \right\}  _{r\in f\left(  U\right)  }$. Then he moved to the QM case
where the universe set, e.g., $U=\left\{  u_{1},...,u_{n}\right\}  $, or
"aggregate of $n$ states has to be replaced by an $n$-dimensional Euclidean
vector space" \cite[p. 256]{weyl:phil}. The appropriate notion of a vector
space partition or "grating" is a "splitting of the total vector space into
mutually orthogonal subspaces" so that "each vector $\overrightarrow{x}$
splits into $r$ component vectors lying in the several subspaces" \cite[p.
256]{weyl:phil}, i.e., a direct sum decomposition of the space. After
referring to a partition as a "grating" or "sieve," Weyl notes that
"Measurement means application of a sieve or grating" \cite[p. 259]%
{weyl:phil}, e.g., in QM/sets, the application (i.e., join) of the set-grating
or partition $\left\{  f^{-1}\left(  r\right)  \right\}  _{r\in f\left(
U\right)  }$ to the pure state $\left\{  S\right\}  $ to give the mixed state
$\left\{  f^{-1}\left(  r\right)  \cap S\right\}  _{r\in f\left(  U\right)  }$.

For some mental imagery of measurement, we might think of the grating as a
series of regular-polygonal-shaped holes that might shape an indefinite blob
of dough. In a measurement, the blob of dough falls through one of the
polygonal holes with equal probability and then takes on that shape.%

%TCIMACRO{\FRAME{dtbpF}{2.342in}{2.0374in}{0pt}{}{}{fig2-grating.eps}%
%{\special{ language "Scientific Word";  type "GRAPHIC";
%maintain-aspect-ratio TRUE;  display "USEDEF";  valid_file "F";
%width 2.342in;  height 2.0374in;  depth 0pt;  original-width 3.1731in;
%original-height 2.7571in;  cropleft "0";  croptop "1";  cropright "1";
%cropbottom "0";  filename 'fig2-grating.eps';file-properties "XNPEU";}} }%
%BeginExpansion
\begin{center}
\includegraphics[
height=2.0374in,
width=2.342in
]%
{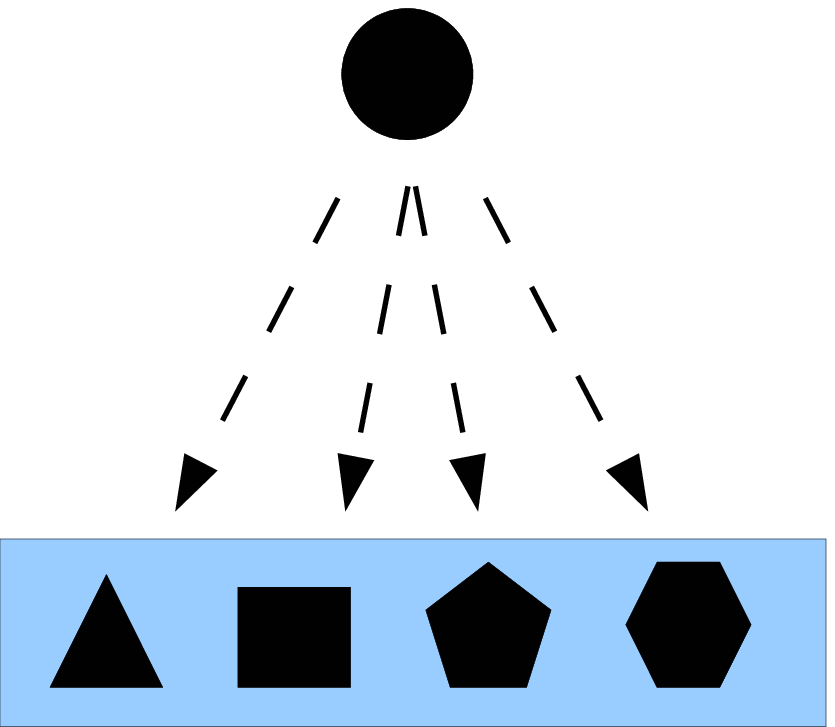}%
\end{center}
%EndExpansion

\begin{center}
Figure 2: Measurement as randomly giving an indefinite blob of dough a
definite polygonal shape.
\end{center}

\subsection{Example of a nondegenerate measurement}

In the simple example illustrated below, we start at the one block or state of
the indiscrete partition or blob which is the completely indistinct entity
$\left\{  a,b,c\right\}  $. A measurement always uses some attribute that
defines an inverse-image partition on $U=\left\{  a,b,c\right\}  $. In the
case at hand, there are "essentially" four possible attributes that could be
used to "measure" the indefinite entity $\left\{  a,b,c\right\}  $ (since
there are four partitions that refine the indiscrete partition in Figure 3).

For an example of a nondegenerate measurement in QM/sets, consider any
attribute $f:U\rightarrow%
%TCIMACRO{\U{211d} }%
%BeginExpansion
\mathbb{R}
%EndExpansion
$ which has the discrete partition as its inverse image (i.e., is injective),
such as the ordinal number of the letter in the alphabet: $f\left(  a\right)
=1$, $f\left(  b\right)  =2$, and $f\left(  c\right)  =3$. This attribute has
three (nonzero) eigenvectors: $f\upharpoonright\left\{  a\right\}  =1\left\{
a\right\}  $, $f\upharpoonright\left\{  b\right\}  =2\left\{  b\right\}  $,
and $f\upharpoonright\left\{  c\right\}  =3\left\{  c\right\}  $ with the
corresponding eigenvalues. The eigenvectors are $\left\{  a\right\}  $,
$\left\{  b\right\}  $, and $\left\{  c\right\}  $, the blocks in the discrete
partition of $U$. The nondegenerate measurement using the observable $f$ acts
on the pure state $U=\left\{  a,b,c\right\}  $ to give the mixed state of the
discrete partition $\mathbf{1}$:

\begin{center}
$U\rightarrow\left\{  U\cap f^{-1}\left(  r\right)  \right\}  _{r=1,2,3}%
=\mathbf{1}$.
\end{center}

\noindent Each such measurement would return an eigenvalue $r$ with the
probability of $\Pr\left(  r|S\right)  =\frac{\left\vert f^{-1}\left(
r\right)  \cap S\right\vert }{\left\vert S\right\vert }=\frac{1}{3}$ for $r\in
f\left(  U\right)  =\left\{  1,2,3\right\}  $.

A projective measurement makes distinctions in the measured state that are
sufficient to induce the "quantum jump" or projection to the eigenvector
associated with the observed eigenvalue. If the observed eigenvalue was $3$,
then the state $\left\{  a,b,c\right\}  $ projects to $f^{-1}\left(  3\right)
\cap\left\{  a,b,c\right\}  =\left\{  c\right\}  \cap\left\{  a,b,c\right\}
=\left\{  c\right\}  $ as pictured below.%

%TCIMACRO{\FRAME{dtbpF}{2.2349in}{1.457in}{0in}{}{}{fig3-measpathnd.eps}%
%{\special{ language "Scientific Word";  type "GRAPHIC";
%maintain-aspect-ratio TRUE;  display "USEDEF";  valid_file "F";
%width 2.2349in;  height 1.457in;  depth 0in;  original-width 2.1295in;
%original-height 1.3782in;  cropleft "0";  croptop "1";  cropright "1";
%cropbottom "0";  filename 'fig3-measpathnd.eps';file-properties "XNPEU";}} }%
%BeginExpansion
\begin{center}
\includegraphics[
height=1.457in,
width=2.2349in
]%
{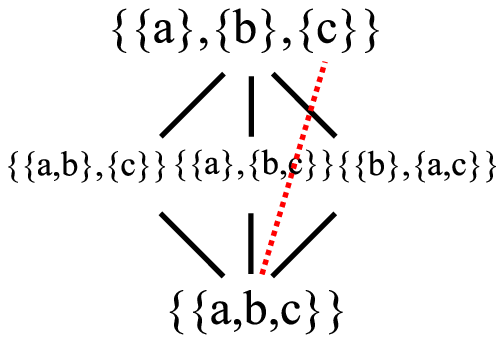}%
\end{center}
%EndExpansion

\begin{center}
Figure 3: Nondegenerate measurement and resulting "quantum jump"
\end{center}

It might be emphasized that this is a state reduction from the single
indefinite state $\left\{  a,b,c\right\}  $ to the single definite state
$\left\{  c\right\}  $, not a subjective removal of ignorance as if the state
had all along been $\left\{  c\right\}  $.

\subsection{Example of a degenerate measurement}

For an example of a degenerate measurement, we choose an attribute with a
non-discrete inverse-image partition such as the partition $\pi=\left\{
\left\{  a\right\}  ,\left\{  b,c\right\}  \right\}  $. Hence the attribute
could just be the characteristic function $\chi_{\left\{  b,c\right\}  }$ with
the two eigenspaces $\wp(\left\{  a\right\}  )$ and $\wp(\left\{  b,c\right\}
)$ and the two eigenvalues $0$ and $1$ respectively. Since the eigenspace
$\wp\left(  \chi_{\left\{  b,c\right\}  }^{-1}\left(  1\right)  \right)
=\wp\left(  \left\{  b,c\right\}  \right)  $ is not one dimensional, the
eigenvalue of $1$ is a QM/sets-version of a \textit{degenerate} eigenvalue.
This attribute $\chi_{\left\{  b,c\right\}  }$ has four (non-zero) eigenvectors:

\begin{center}
$\chi_{\left\{  b,c\right\}  }\upharpoonright\left\{  b,c\right\}  =1\left\{
b,c\right\}  $, $\chi_{\left\{  b,c\right\}  }\upharpoonright\left\{
b\right\}  =1\left\{  b\right\}  $, $\chi_{\left\{  b,c\right\}
}\upharpoonright\left\{  c\right\}  =1\left\{  c\right\}  $, and
$\chi_{\left\{  b,c\right\}  }\upharpoonright\left\{  a\right\}  =0\left\{
a\right\}  $.
\end{center}

The "measuring apparatus" makes distinctions by joining the attribute
inverse-image partition

\begin{center}
$\chi_{\left\{  b,c\right\}  }^{-1}=\left\{  \chi_{\left\{  b,c\right\}
}^{-1}\left(  1\right)  ,\chi_{\left\{  b,c\right\}  }^{-1}\left(  0\right)
\right\}  =\left\{  \left\{  b,c\right\}  ,\left\{  a\right\}  \right\}  $
\end{center}

\noindent with the pure state representing the indefinite entity $U=\left\{
a,b,c\right\}  $. The action on the pure state is:

\begin{center}
$U\rightarrow\left\{  U\right\}  \vee\chi_{\left\{  b,c\right\}  }^{-1}%
=\chi_{\left\{  b,c\right\}  }^{-1}=\left\{  \left\{  b,c\right\}  ,\left\{
a\right\}  \right\}  $.
\end{center}

\noindent The measurement of that attribute returns one of the eigenvalues
with the probabilities:

\begin{center}
$\Pr(0|U)=\frac{\left\vert \left\{  a\right\}  \cap\left\{  a,b,c\right\}
\right\vert }{\left\vert \left\{  a,b,c\right\}  \right\vert }=\frac{1}{3}$
and $\Pr\left(  1|U\right)  =\frac{\left\vert \left\{  b,c\right\}
\cap\left\{  a,b,c\right\}  \right\vert }{\left\vert \left\{  a,b,c\right\}
\right\vert }=\frac{2}{3}$.
\end{center}

\noindent Suppose it returns the eigenvalue $1$. Then the indefinite entity
$\left\{  a,b,c\right\}  $ reduces to the projected eigenstate $\chi_{\left\{
b,c\right\}  }^{-1}\left(  1\right)  \cap\left\{  a,b,c\right\}  =\left\{
b,c\right\}  $ for that eigenvalue \cite[p. 221]{cohen-t:QM1}.

Since this is a degenerate result (i.e., the eigenspace $\wp\left(
\chi_{\left\{  b,c\right\}  }^{-1}\left(  1\right)  \right)  =\wp\left(
\left\{  b,c\right\}  \right)  $ doesn't have dimension one), another
measurement is needed to make more distinctions. Measurements by attributes,
such as $\chi_{\left\{  a,b\right\}  }$ or $\chi_{\left\{  a,c\right\}  }$,
that give either of the other two partitions, $\left\{  \left\{  a,b\right\}
,\{c\right\}  \}$ or $\left\{  \left\{  b\right\}  ,\left\{  a,c\right\}
\right\}  $ as inverse images, would suffice to distinguish $\left\{
b,c\right\}  $ into $\left\{  b\right\}  $ or $\left\{  c\right\}  $. Then
either attribute together with the attribute $\chi_{\left\{  b,c\right\}  }$
would form a \textit{Complete Set of Compatible Attributes} or CSCA (i.e., the
QM/sets-version of a Complete Set of Commuting Operators or CSCO
\cite{dirac:principles}), where \textit{complete} means that the join of the
attributes' inverse-image partitions gives the discrete partition and where
\textit{compatible} means that all the attributes can be taken as defined on
the same set of (simultaneous) basis eigenvectors, e.g., the $U$-basis.

Taking, for example, the other attribute as $\chi_{\left\{  a,b\right\}  }$,
the join of the two attributes' partitions is discrete:

\begin{center}
$\mathbf{\chi}_{\left\{  b,c\right\}  }^{-1}\vee\mathbf{\chi}_{\left\{
a,b\right\}  }^{-1}=\left\{  \left\{  a\right\}  ,\left\{  b,c\right\}
\right\}  \vee\left\{  \left\{  a,b\right\}  ,\{c\right\}  \}=\left\{
\left\{  a\right\}  ,\left\{  b\right\}  ,\left\{  c\right\}  \right\}
=\mathbf{1}$.
\end{center}

\noindent Hence all the eigenstate singletons can be characterized by the
ordered pairs of the eigenvalues of these two attributes: $\left\{  a\right\}
=\left\vert 0,1\right\rangle $, $\left\{  b\right\}  =\left\vert
1,1\right\rangle $, and $\left\{  c\right\}  =\left\vert 1,0\right\rangle $
(using Dirac's ket-notation to give the ordered pairs and listing the
eigenvalues of $\chi_{\left\{  b,c\right\}  }$ first on the left).

The second projective measurement of the indefinite entity $\left\{
b,c\right\}  $ using the attribute $\chi_{\left\{  a,b\right\}  }$ with the
inverse-image partition $\chi_{\left\{  a,b\right\}  }^{-1}=\left\{  \left\{
a,b\right\}  ,\{c\right\}  \}$ would have the pure-to-mixed state action:

\begin{center}
$\left\{  b,c\right\}  \rightarrow\left\{  \left\{  b,c\right\}  \cap
\chi_{\left\{  a,b\right\}  }(1),\left\{  b,c\right\}  \cap\chi_{\left\{
a,b\right\}  }\left(  0\right)  \right\}  =\left\{  \left\{  b\right\}
,\left\{  c\right\}  \right\}  $.
\end{center}

The distinction-making measurement would cause the indefinite entity $\left\{
b,c\right\}  $ to turn into one of the definite entities of $\left\{
b\right\}  $ or $\left\{  c\right\}  $ with the probabilities:

\begin{center}
$\Pr\left(  1|\left\{  b,c\right\}  \right)  =\frac{\left\vert \left\{
a,b\right\}  \cap\left\{  b,c\right\}  \right\vert }{\left\vert \left\{
b,c\right\}  \right\vert }=\frac{1}{2}$ and $\Pr\left(  0|\left\{
b,c\right\}  \right)  =\frac{\left\vert \left\{  c\right\}  \cap\left\{
b,c\right\}  \right\vert }{\left\vert \left\{  b,c\right\}  \right\vert
}=\frac{1}{2}$.
\end{center}

\noindent If the measured eigenvalue is $0$, then the state $\left\{
b,c\right\}  $ projects to $\chi_{\left\{  a,b\right\}  }^{-1}\left(
0\right)  \cap\left\{  b,c\right\}  =\left\{  c\right\}  $ as pictured below.%

%TCIMACRO{\FRAME{dtbpF}{2.2349in}{1.457in}{0in}{}{}{fig4-measpath.eps}%
%{\special{ language "Scientific Word";  type "GRAPHIC";
%maintain-aspect-ratio TRUE;  display "USEDEF";  valid_file "F";
%width 2.2349in;  height 1.457in;  depth 0in;  original-width 2.1295in;
%original-height 1.3782in;  cropleft "0";  croptop "1";  cropright "1";
%cropbottom "0";  filename 'fig4-measpath.eps';file-properties "XNPEU";}} }%
%BeginExpansion
\begin{center}
\includegraphics[
height=1.457in,
width=2.2349in
]%
{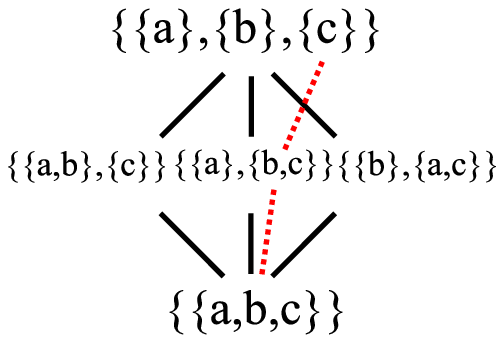}%
\end{center}
%EndExpansion

\begin{center}
Figure 4: Degenerate measurement
\end{center}

\noindent The two projective measurements of $\left\{  a,b,c\right\}  $ using
the complete set of compatible (e.g., both defined on $U$) attributes
$\chi_{\left\{  b,c\right\}  }$ and $\chi_{\left\{  a,b\right\}  }$ produced
the respective eigenvalues $1$ and $0$ so the resulting eigenstate was
characterized by the eigenket $\left\vert 1,0\right\rangle =\{c\}$.

Again, this is all analogous to standard Dirac-von-Neumann quantum mechanics.

\subsection{Measurement using density matrices}

The previous treatment of the role of partitions in measurement can be
restated using density matrices \cite[p. 98]{nielsen-chuang:bible} over the
reals. Given a partition $\pi=\left\{  B\right\}  $ on $U=\left\{
u_{1},...,u_{n}\right\}  $, the blocks $B\in\pi$ can be thought of as
(nonoverlapping or "orthogonal") "pure states" where the "state" $B$ occurs
with the probability $p_{B}=\frac{\left\vert B\right\vert }{\left\vert
U\right\vert }$. Then we can mimic the usual procedure for forming the density
matrix $\rho\left(  \pi\right)  $ for the "orthogonal pure states" $B$ with
the probabilities $p_{B}$. The "pure state" $B$ normalized in the reals to
length $1$ is represented by the column vector $\left\vert B\right\rangle
_{1}=\frac{1}{\sqrt{\left\vert B\right\vert }}\left[  \chi_{B}\left(
u_{1}\right)  ,...,\chi_{B}\left(  u_{n}\right)  \right]  ^{t}$(where $\left[
{}\right]  ^{t}$ indicates the transpose). Then the \textit{density matrix
}$\rho\left(  B\right)  $\textit{\ for the pure state }$B\subseteq U$ is then
(calculating in the reals):

\begin{center}
$\rho\left(  B\right)  =\left\vert B\right\rangle _{1}\left(  \left\vert
B\right\rangle _{1}\right)  ^{t}=\frac{1}{\left\vert B\right\vert }%
\begin{bmatrix}
\chi_{B}\left(  u_{1}\right) \\
\chi_{B}\left(  u_{2}\right) \\
\vdots\\
\chi_{B}\left(  u_{n}\right)
\end{bmatrix}
\left[  \chi_{B}\left(  u_{1}\right)  ,...,\chi_{B}\left(  u_{n}\right)
\right]  $

$=\frac{1}{\left\vert B\right\vert }%
\begin{bmatrix}
\chi_{B}\left(  u_{1}\right)  & \chi_{B}\left(  u_{1}\right)  \chi_{B}\left(
u_{2}\right)  & \cdots & \chi_{B}\left(  u_{1}\right)  \chi_{B}\left(
u_{n}\right) \\
\chi_{B}\left(  u_{2}\right)  \chi_{B}\left(  u_{1}\right)  & \chi_{B}\left(
u_{2}\right)  & \cdots & \chi_{B}\left(  u_{2}\right)  \chi_{B}\left(
u_{n}\right) \\
\vdots & \vdots & \ddots & \vdots\\
\chi_{B}\left(  u_{n}\right)  \chi_{B}\left(  u_{1}\right)  & \chi_{B}\left(
u_{n}\right)  \chi_{B}\left(  u_{2}\right)  & \cdots & \chi_{B}\left(
u_{n}\right)
\end{bmatrix}
$.
\end{center}

For instance if $U=\left\{  u_{1},u_{2},u_{3}\right\}  $, then for the blocks
in the partition $\pi=\left\{  \left\{  u_{1},u_{2}\right\}  ,\left\{
u_{3}\right\}  \right\}  $:

\begin{center}
$\rho\left(  \left\{  u_{1},u_{2}\right\}  \right)  =%
\begin{bmatrix}
\frac{1}{2} & \frac{1}{2} & 0\\
\frac{1}{2} & \frac{1}{2} & 0\\
0 & 0 & 0
\end{bmatrix}
$ and $\rho\left(  \left\{  u_{3}\right\}  \right)  =%
\begin{bmatrix}
0 & 0 & 0\\
0 & 0 & 0\\
0 & 0 & 1
\end{bmatrix}
$.
\end{center}

\noindent Then the "mixed state" \textit{density matrix }$\rho\left(
\pi\right)  $\textit{\ of the partition} $\pi$ is the weighted sum:

\begin{center}
$\rho\left(  \pi\right)  =\sum_{B\in\pi}p_{B}\rho\left(  B\right)  $.
\end{center}

In the example, this is:

\begin{center}
$\rho\left(  \pi\right)  =\frac{2}{3}%
\begin{bmatrix}
\frac{1}{2} & \frac{1}{2} & 0\\
\frac{1}{2} & \frac{1}{2} & 0\\
0 & 0 & 0
\end{bmatrix}
+\frac{1}{3}%
\begin{bmatrix}
0 & 0 & 0\\
0 & 0 & 0\\
0 & 0 & 1
\end{bmatrix}
=%
\begin{bmatrix}
\frac{1}{3} & \frac{1}{3} & 0\\
\frac{1}{3} & \frac{1}{3} & 0\\
0 & 0 & \frac{1}{3}%
\end{bmatrix}
$.
\end{center}

\noindent While this construction mimics the usual construction of the density
matrix for orthogonal pure states, the remarkable thing is that the entries
have a direct interpretation in terms of the dits and indits of the partition
$\pi$:

\begin{center}
$\rho_{jk}\left(  \pi\right)  =\left\{
\begin{array}
[c]{c}%
\frac{1}{\left\vert U\right\vert }\text{ if }\left(  u_{j},u_{k}\right)
\in\operatorname*{indit}\left(  \pi\right) \\
0\text{ if }\left(  u_{j},u_{k}\right)  \notin\operatorname*{indit}\left(
\pi\right)
\end{array}
\right.  $.
\end{center}

\noindent All the entries are real "amplitudes" whose squares are the two-draw
probabilities of drawing a pair of elements from $U$ (with replacement) that
is an indistinction of $\pi$. As in the quantum case, the non-zero entries of
the density matrix $\rho_{jk}\left(  \pi\right)  =\sqrt{\frac{1}{\left\vert
U\right\vert }\frac{1}{\left\vert U\right\vert }}=\frac{1}{\left\vert
U\right\vert }$ are the "coherences" \cite[p. 302]{cohen-t:QM1} which indicate
that $u_{j}$ and $u_{k}$ "cohere" together in a block or "pure state" of the
partition, i.e., $\left(  u_{j},u_{k}\right)  \in\operatorname*{indit}\left(
\pi\right)  $. Since the ordered pairs $\left(  u_{j},u_{j}\right)  $ in the
diagonal $\Delta\subseteq U\times U$ are always indits of any partition, the
diagonal entries in $\rho\left(  \pi\right)  $ are always $\frac{1}{\left\vert
U\right\vert }$.

Combinatorial theory gives another way to define the density matrix of a
partition. A binary relation $R\subseteq U\times U$ on $U=\left\{
u_{1},...,u_{n}\right\}  $ can be represented by an $n\times n$
\textit{incidence matrix} $I(R)$ where

\begin{center}
$I\left(  R\right)  _{ij}=\left\{
\begin{array}
[c]{c}%
1\text{ if }\left(  u_{i},u_{j}\right)  \in R\\
0\text{ if }\left(  u_{i},u_{j}\right)  \notin R\text{.}%
\end{array}
\right.  $
\end{center}

\noindent Taking $R$ as the equivalence relation $\operatorname*{indit}\left(
\pi\right)  $ associated with a partition $\pi$, the density matrix
$\rho\left(  \pi\right)  $ defined above is just the incidence matrix
$I\left(  \operatorname*{indit}\left(  \pi\right)  \right)  $ normalized to be
of trace $1$ (sum of diagonal entries is $1$):

\begin{center}
$\rho\left(  \pi\right)  =\frac{1}{\left\vert U\right\vert }I\left(
\operatorname*{indit}\left(  \pi\right)  \right)  $.
\end{center}

If the subsets $T\in\wp\left(  U\right)  $ are represented by the $n$-ary
column vectors $\left[  \chi_{T}\left(  u_{1}\right)  ,...,\chi_{T}\left(
u_{n}\right)  \right]  ^{t}$, then the action of the projection operator
$B\cap():\wp\left(  U\right)  \rightarrow\wp\left(  U\right)  $ is represented
by the $n\times n$ diagonal matrix $P_{B}$ where the diagonal entries are:

\begin{center}
$\left(  P_{B}\right)  _{ii}=\left\{
\begin{array}
[c]{c}%
1\text{ if }u_{i}\in B\\
0\text{ if }u_{i}\notin B
\end{array}
\right.  =\chi_{B}\left(  u_{i}\right)  $
\end{center}

\noindent which is idempotent, $P_{B}^{2}=P_{B}$, and symmetric, $P_{B}%
^{t}=P_{B}$. For any state $S\in\wp\left(  U\right)  $, the trace (sum of
diagonal entries) of $P_{B}\rho\left(  S\right)  $ is:

\begin{center}
$\operatorname*{tr}\left[  P_{B}\rho\left(  S\right)  \right]  =\frac
{1}{\left\vert S\right\vert }\sum_{i=1}^{n}\chi_{S}\left(  u_{i}\right)
\chi_{B}\left(  u_{i}\right)  =\frac{\left\vert B\cap S\right\vert
}{\left\vert S\right\vert }=\Pr\left(  B|S\right)  $
\end{center}

\noindent so given $f:U\rightarrow%
%TCIMACRO{\U{211d} }%
%BeginExpansion
\mathbb{R}
%EndExpansion
$,

\begin{center}
$\Pr\left(  r|S\right)  =\frac{\left\vert f^{-1}\left(  r\right)  \cap
S\right\vert }{\left\vert S\right\vert }=\operatorname*{tr}\left[
P_{f^{-1}\left(  r\right)  }\rho\left(  S\right)  \right]  $.
\end{center}

We saw previously how the action of a measurement in QM/sets could be
described using the partition join operation. The join $\pi\vee\sigma$ of the
partitions $\pi=\left\{  B\right\}  $ and $\sigma=\left\{  C\right\}  $ could
be seen as the result $%
%TCIMACRO{\tbigcup \nolimits_{C\in\sigma}}%
%BeginExpansion
{\textstyle\bigcup\nolimits_{C\in\sigma}}
%EndExpansion
\left\{  C\cap B\neq\emptyset:B\in\pi\right\}  $ of the projection operators
$C\cap()$ acting on the $B\in\pi$ for all $C\in\sigma$. Substituting the
normalized $\left\vert B\right\rangle _{1}$ for $B$ with the density matrix
$\rho\left(  B\right)  =\left\vert B\right\rangle _{1}\left(  \left\vert
B\right\rangle _{1}\right)  ^{t}$ and the matrix projection operators $P_{C}$
for $C\cap()$, the application of $P_{C}$ to $\left\vert B\right\rangle _{1}$
yields the density matrix:

\begin{center}
$\left(  P_{C}\left\vert B\right\rangle _{1}\right)  \left(  P_{C}\left\vert
B\right\rangle _{1}\right)  ^{t}=P_{C}\left\vert B\right\rangle _{1}\left(
\left\vert B\right\rangle _{1}\right)  ^{t}P_{C}^{t}=P_{C}\rho\left(
B\right)  P_{C}$.
\end{center}

\noindent Summing with the probability weights gives: $\sum_{B\in\pi}%
p_{B}P_{C}\rho\left(  B\right)  P_{C}=P_{C}\rho\left(  \pi\right)  P_{C}$ and
then summing over the different projection operators gives: $\sum_{C\in\sigma
}P_{C}\rho\left(  \pi\right)  P_{C}$. A little calculation then shows that
this is exactly the \textit{density matrix of the partition join}:

\begin{center}
$\sum_{C\in\sigma}P_{C}\rho\left(  \pi\right)  P_{C}=\rho\left(  \pi\vee
\sigma\right)  $.

Density matrix version of the partition join
\end{center}

We are modeling, using density matrices, the QM/sets projective measurement of
an attribute $f:U\rightarrow%
%TCIMACRO{\U{211d} }%
%BeginExpansion
\mathbb{R}
%EndExpansion
$ starting with a pure state $S$. Then measurement converts the pure state
$\left\vert S\right\rangle $ to one of the states $\left\vert f^{-1}\left(
r\right)  \cap S\right\rangle $ with the probability $\frac{|f^{-1}\left(
r\right)  \cap S|}{\left\vert S\right\vert }$. In the previous example of a
(degenerate) measurement with $U=\left\{  a,b,c\right\}  $ and $f=\chi
_{\left\{  b,c\right\}  }$, then the measurement, in terms of partitions, had
the effect of making distinctions on the partition $\left\{  U\right\}  $ by
the partition $\left\{  f^{-1}\right\}  =\left\{  \chi_{\left\{  b,c\right\}
}^{-1}\right\}  $using the join operation:

\begin{center}
$\left\{  U\right\}  \rightarrow\left\{  U\right\}  \vee\left\{
\chi_{\left\{  b,c\right\}  }^{-1}\right\}  =\left\{  \left\{  b,c\right\}
,\left\{  a\right\}  \right\}  $.
\end{center}

\noindent The mixed state $\left\{  \left\{  b,c\right\}  ,\left\{  a\right\}
\right\}  $ has the projected outcomes $\chi_{\left\{  b,c\right\}  }%
^{-1}\left(  1\right)  \cap U=\left\{  b,c\right\}  $ and $\chi_{\left\{
b,c\right\}  }^{-1}\left(  0\right)  \cap U=\left\{  a\right\}  $ which occur
with the probabilities $\Pr\left(  1|U\right)  =\left\vert \chi_{\left\{
b,c\right\}  }^{-1}\left(  1\right)  \cap U\right\vert /\left\vert
U\right\vert =2/3$ and $\Pr\left(  0|U\right)  =\left\vert \chi_{\left\{
b,c\right\}  }^{-1}\left(  0\right)  \cap U\right\vert /\left\vert
U\right\vert =1/3$.

We now have the density matrix version of the partition join operation, so in
the general case of starting with the pure state $S$, we might take the
starting partition on $U$ as $\pi=\left\{  S,S^{c}\right\}  $ and then take
the measurement join with $\sigma=\left\{  f^{-1}\right\}  =\left\{
f^{-1}\left(  r\right)  \right\}  _{r\in f\left(  U\right)  }$ which yields
the density matrix (using linearity):

\begin{center}
$\rho\left(  \left\{  S,S^{c}\right\}  \vee\left\{  f^{-1}\right\}  \right)
=\sum_{r\in f\left(  U\right)  }P_{f^{-1}\left(  r\right)  }\rho\left(
\left\{  S,S^{c}\right\}  \right)  P_{f^{-1}\left(  r\right)  }$

$=p_{S}\sum_{r\in f\left(  U\right)  }P_{f^{-1}\left(  r\right)  }\rho\left(
S\right)  P_{f^{-1}\left(  r\right)  }+p_{S^{c}}\sum_{r\in f\left(  U\right)
}P_{f^{-1}\left(  r\right)  }\rho\left(  S^{c}\right)  P_{f^{-1}\left(
r\right)  }$.
\end{center}

\noindent Thus starting with the pure state density matrix $\rho\left(
S\right)  =\left\vert S\right\rangle _{1}\left(  \left\vert S\right\rangle
_{1}\right)  ^{t}$, the action of the measurement given by the partition join
(ignoring the action on the complement $S^{c}$) is to create the mixed state
$\hat{\rho}\left(  S\right)  $:

\begin{center}
\fbox{$\rho\left(  S\right)  \longrightarrow\hat{\rho}\left(  S\right)
=\sum_{r\in f\left(  U\right)  }P_{f^{-1}\left(  r\right)  }\rho\left(
S\right)  P_{f^{-1}\left(  r\right)  }$}

Action of measurement of attribute $f$ on the pure state density matrix
$\rho\left(  S\right)  $.
\end{center}

\noindent In that mixed state, the projected state $\left\vert f^{-1}\left(
r\right)  \cap S\right\rangle $ occurs with the probability
$\operatorname*{tr}[P_{f^{-1}\left(  r\right)  }\rho\left(  S\right)
]=\frac{|f^{-1}\left(  r\right)  \cap S|}{\left\vert S\right\vert }=\Pr\left(
r|S\right)  $.

In full QM, the projective measurement using a Hermitian observable operator
$L$ with the spectral decomposition $L=\sum_{i=1}^{m}\lambda_{i}P_{i}$ of a
normalized pure state $\left\vert \psi\right\rangle $ results in the state
$P_{i}\left\vert \psi\right\rangle $ with the probability $p_{i}%
=\operatorname*{tr}\left[  P_{i}\rho\left(  \psi\right)  \right]  =\Pr\left(
\lambda_{i}|\psi\right)  $ where $\rho\left(  \psi\right)  =\left\vert
\psi\right\rangle \left\langle \psi\right\vert $. The projected resultant
state $P_{i}\left\vert \psi\right\rangle $ has the density matrix $\frac
{P_{i}\left\vert \psi\right\rangle \left\langle \psi\right\vert P_{i}%
}{\operatorname*{tr}\left[  P_{i}\rho\left(  \psi\right)  \right]  }%
=\frac{P_{i}\rho\left(  \psi\right)  P_{i}}{\operatorname*{tr}\left[
P_{i}\rho\left(  \psi\right)  \right]  }$ so the mixed state describing the
probabilistic results of the measurement is \cite[p. 101]%
{nielsen-chuang:bible}:

\begin{center}
$\hat{\rho}\left(  \psi\right)  =\sum_{i}p_{i}\frac{P_{i}\rho\left(
\psi\right)  P_{i}}{\operatorname*{tr}\left[  P_{i}\rho\left(  \psi\right)
\right]  }=\sum_{i}\operatorname*{tr}\left[  P_{i}\rho\left(  \psi\right)
\right]  \frac{P_{i}\rho\left(  \psi\right)  P_{i}}{\operatorname*{tr}\left[
P_{i}\rho\left(  \psi\right)  \right]  }=\sum_{i}P_{i}\rho\left(  \psi\right)
P_{i}$.
\end{center}

Thus we see how the density matrix treatment of measurement in QM/sets is just
a sets-version of the density matrix treatment of projective measurement in
standard Dirac-von-Neumann QM. And we have the additional
philosophically-relevant information that the measurement is described by the
distinction-creating partition join operation in QM/sets--which confirms the
observation in QM that the essence of measurement is distinguishing the
alternative possible states.

For instance, Richard Feynman always emphasized the importance of distinctions
as characterizing what amounts to a "measurement."

\begin{quotation}
\noindent If you could, in principle, distinguish the alternative final states
(even though you do not bother to do so), the total, final probability is
obtained by calculating the probability for each state (not the amplitude) and
then adding them together. If you cannot distinguish the final states even in
principle, then the probability amplitudes must be summed before taking the
absolute square to find the actual probability.\cite[p. 3.9]{feynman:vol3}
\end{quotation}

\subsection{Quantum dynamics and the two-slit experiment in QM/sets}

To illustrate a two-slit experiment in quantum mechanics over sets, we need to
introduce some "dynamics." In quantum mechanics, the no-distinctions
requirement is that the linear transformation has to preserve the degree of
indistinctness $\left\langle \psi|\varphi\right\rangle $, i.e., that it
preserved the inner product. Where two normalized states are fully distinct if
$\left\langle \psi|\varphi\right\rangle =0$ and fully indistinct if
$\left\langle \psi|\varphi\right\rangle =1$, it is also sufficient to just
require that full distinctness and indistinctness be preserved since that
would imply orthonormal bases are preserved and that is equivalent to being
unitary. In QM/sets, we have no inner product but the idea of a linear
transformation $A:%
%TCIMACRO{\U{2124} }%
%BeginExpansion
\mathbb{Z}
%EndExpansion
_{2}^{n}\rightarrow%
%TCIMACRO{\U{2124} }%
%BeginExpansion
\mathbb{Z}
%EndExpansion
_{2}^{n}$ preserving distinctness would simply mean being non-singular. The
condition analogous to preserving inner product is $\left\langle
S|_{U}T\right\rangle =\left\langle A\left(  S\right)  |_{A\left(  U\right)
}A\left(  T\right)  \right\rangle $ where $A\left(  U\right)  =U^{\prime}$ is
defined by $A\left(  \left\{  u\right\}  \right)  =\left\{  u^{\prime
}\right\}  $. For non-singular $A$, the image $A\left(  U\right)  $ of the
$U$-basis is a basis, i.e., the $U^{\prime}$-basis, and the
"bracket-preserving" condition holds since $\left\vert S\cap T\right\vert
=\left\vert A\left(  S\right)  \cap A\left(  T\right)  \right\vert $ for
$A\left(  S\right)  ,A\left(  T\right)  \subseteq A\left(  U\right)
=U^{\prime}$. Hence the QM/sets analogue of the unitary dynamics of full QM is
"non-singular dynamics," i.e., the change-of-state matrix is
non-singular.\footnote{In Schumacher and Westmoreland's modal quantum theory
\cite{schum:modal}, they also take the dynamics to be any non-singular linear
transformation.}

Consider the dynamics given in terms of the $U$-basis where: $\left\{
a\right\}  \rightarrow\left\{  a,b\right\}  $; $\left\{  b\right\}
\rightarrow\left\{  a,b,c\right\}  $; and $\left\{  c\right\}  \rightarrow
\left\{  b,c\right\}  $ in one time period. This is represented by the
non-singular one-period change of state matrix:

\begin{center}
$A=%
\begin{bmatrix}
1 & 1 & 0\\
1 & 1 & 1\\
0 & 1 & 1
\end{bmatrix}
$.
\end{center}

The seven nonzero vectors in the vector space are divided by this "dynamics"
into a $4$ -orbit: $\left\{  a\right\}  \rightarrow\left\{  a,b\right\}
\rightarrow\left\{  c\right\}  \rightarrow\left\{  b,c\right\}  \rightarrow
\left\{  a\right\}  $, a $2$-orbit: $\left\{  b\right\}  \rightarrow\left\{
a,b,c\right\}  \rightarrow\left\{  b\right\}  $, and a $1$-orbit: $\left\{
a,c\right\}  \rightarrow\left\{  a,c\right\}  $.

If we take the $U$-basis vectors as "vertical position" eigenstates, we can
device a QM/sets version of the "two-slit experiment" which models "all of the
mystery of quantum mechanics" \cite[p. 130]{fey-phylaw}. Taking $a$, $b$, and
$c$ as three vertical positions, we have a vertical diaphragm with slits at
$a$ and $c$. Then there is a screen or wall to the right of the slits so that
a "particle" will travel from the diaphragm to the wall in one time period
according to the $A$-dynamics.%

%TCIMACRO{\FRAME{dtbpF}{2.5139in}{1.8182in}{0in}{}{}{fig5-two-slit-setup.eps}%
%{\special{ language "Scientific Word";  type "GRAPHIC";
%maintain-aspect-ratio TRUE;  display "USEDEF";  valid_file "F";
%width 2.5139in;  height 1.8182in;  depth 0in;  original-width 2.3993in;
%original-height 1.7277in;  cropleft "0";  croptop "1";  cropright "1";
%cropbottom "0";  filename 'fig5-two-slit-setup.eps';file-properties "XNPEU";}}
%}%
%BeginExpansion
\begin{center}
\includegraphics[
height=1.8182in,
width=2.5139in
]%
{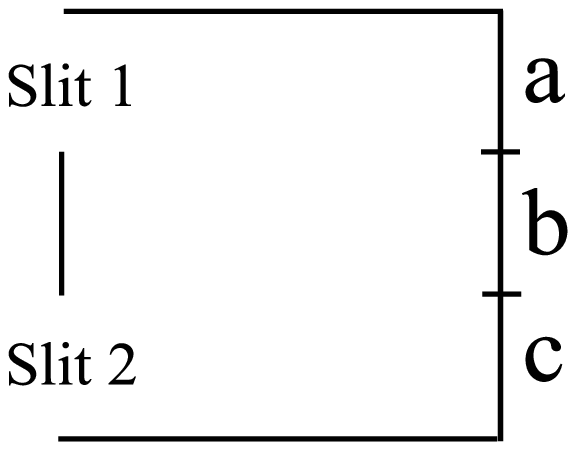}%
\end{center}
%EndExpansion

\begin{center}
Figure 5: Two-slit setup
\end{center}

We start with or "prepare" the state of a particle being at the slits in the
indefinite position state $\left\{  a,c\right\}  $. Then there are two cases.

\textbf{First case of distinctions at slits}: The first case is where we
measure the $U$-state at the slits and then let the resultant position
eigenstate evolve by the $A$-dynamics to hit the wall at the right where the
position is measured again. The probability that the particle is at slit 1 or
at slit 2 is:

\begin{center}
$\Pr\left(  \left\{  a\right\}  \text{ measured at slits }|\left\{
a,c\right\}  \text{ at slits}\right)  =\frac{\left\langle \left\{  a\right\}
|_{U}\left\{  a,c\right\}  \right\rangle ^{2}}{\left\Vert \left\{
a,c\right\}  \right\Vert _{U}^{2}}=\frac{\left\vert \left\{  a\right\}
\cap\left\{  a,c\right\}  \right\vert }{\left\vert \left\{  a,c\right\}
\right\vert }=\frac{1}{2}$;

$\Pr\left(  \left\{  c\right\}  \text{ measured at slits }|\left\{
a,c\right\}  \text{ at slits}\right)  =\frac{\left\langle \left\{  c\right\}
|_{U}\left\{  a,c\right\}  \right\rangle ^{2}}{\left\Vert \left\{
a,c\right\}  \right\Vert _{U}^{2}}=\frac{\left\vert \left\{  c\right\}
\cap\left\{  a,c\right\}  \right\vert }{\left\vert \left\{  a,c\right\}
\right\vert }=\frac{1}{2}$.
\end{center}

If the particle was at slit 1, i.e., was in eigenstate $\left\{  a\right\}  $,
then it evolves in one time period by the $A$-dynamics to $\left\{
a,b\right\}  $ where the position measurements yield the probabilities of
being at $a$ or at $b$ as:%

\begin{align*}
\Pr\left(  \left\{  a\right\}  \text{ measured at wall }|\left\{  a,b\right\}
\text{ at wall}\right)   &  =\frac{\left\langle \left\{  a\right\}
|_{U}\left\{  a,b\right\}  \right\rangle ^{2}}{\left\Vert \left\{
a,b\right\}  \right\Vert _{U}^{2}}=\frac{\left\vert \left\{  a\right\}
\cap\left\{  a,b\right\}  \right\vert }{\left\vert \left\{  a,b\right\}
\right\vert }=\frac{1}{2}\\
\Pr\left(  \left\{  b\right\}  \text{ measured at wall }|\left\{  a,b\right\}
\text{ at wall}\right)   &  =\frac{\left\langle \left\{  b\right\}
|_{U}\left\{  a,b\right\}  \right\rangle ^{2}}{\left\Vert \left\{
a,b\right\}  \right\Vert _{U}^{2}}=\frac{\left\vert \left\{  b\right\}
\cap\left\{  a,b\right\}  \right\vert }{\left\vert \left\{  a,b\right\}
\right\vert }=\frac{1}{2}\text{.}%
\end{align*}

If on the other hand the particle was found in the first measurement to be at
slit 2, i.e., was in eigenstate $\left\{  c\right\}  $, then it evolved in one
time period by the $A$-dynamics to $\left\{  b,c\right\}  $ where the position
measurements yield the probabilities of being at $b$ or at $c$ as:

\begin{center}
$\Pr\left(  \left\{  b\right\}  \text{ measured at wall }|\left\{
b,c\right\}  \text{ at wall}\right)  =\frac{\left\vert \left\{  b\right\}
\cap\left\{  b,c\right\}  \right\vert }{\left\vert \left\{  b,c\right\}
\right\vert }=\frac{1}{2}$

$\Pr\left(  \left\{  c\right\}  \text{ measured at wall }|\left\{
b,c\right\}  \text{ at wall}\right)  =\frac{\left\vert \left\{  c\right\}
\cap\left\{  b,c\right\}  \right\vert }{\left\vert \left\{  b,c\right\}
\right\vert }=\frac{1}{2}$.
\end{center}

Hence we can use the laws of probability theory to compute the probabilities
of the particle being measured at the three positions on the wall at the right
if it starts at the slits in the superposition state $\left\{  a,c\right\}  $
\textit{and} the measurements were made at the slits:

\begin{center}%
\begin{tabular}
[c]{l}%
$\Pr(\left\{  a\right\}  $ measured at wall $|\left\{  a,c\right\}  $ at
slits$)=\frac{1}{2}\frac{1}{2}=\frac{1}{4}$;\\
$\Pr(\left\{  b\right\}  $ measured at wall $|\left\{  a,c\right\}  $ at
slits$)=\frac{1}{2}\frac{1}{2}+\frac{1}{2}\frac{1}{2}=\frac{1}{2}$;\\
$\Pr(\left\{  c\right\}  $ measured at wall $|\left\{  a,c\right\}  $ at
slits$)=\frac{1}{2}\frac{1}{2}=\frac{1}{4}$.
\end{tabular}
%

%TCIMACRO{\FRAME{dtbpF}{2.5139in}{1.8032in}{0in}{}{}{fig6-case1-two-slit.eps}%
%{\special{ language "Scientific Word";  type "GRAPHIC";
%maintain-aspect-ratio TRUE;  display "USEDEF";  valid_file "F";
%width 2.5139in;  height 1.8032in;  depth 0in;  original-width 2.3993in;
%original-height 1.7136in;  cropleft "0";  croptop "1";  cropright "1";
%cropbottom "0";  filename 'fig6-case1-two-slit.eps';file-properties "XNPEU";}}
%}%
%BeginExpansion
\begin{center}
\includegraphics[
height=1.8032in,
width=2.5139in
]%
{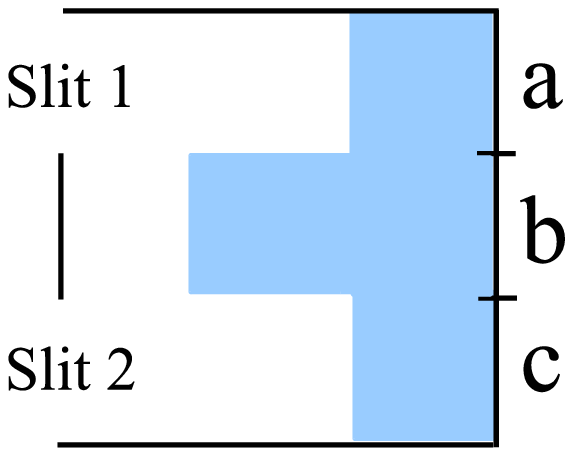}%
\end{center}
%EndExpansion

Figure 6: Final probability distribution with measurements at slits
\end{center}

\textbf{Second case of no distinctions at slits}: The second case is when no
measurements are made at the slits and then the superposition state $\left\{
a,c\right\}  $ evolves by the $A$-dynamics to $\left\{  a,b\right\}  +\left\{
b,c\right\}  =\left\{  a,c\right\}  $ where the superposition at $\left\{
b\right\}  $ cancels out. Then the final probabilities will just be
probabilities of finding $\left\{  a\right\}  $, $\left\{  b\right\}  $, or
$\left\{  c\right\}  $ when the measurement is made only at the wall on the
right is:

\begin{center}%
\begin{tabular}
[c]{l}%
$\Pr(\left\{  a\right\}  $ measured at wall $|\left\{  a,c\right\}  $ at
slits$)=\Pr\left(  \left\{  a\right\}  |\left\{  a,c\right\}  \right)
=\frac{\left\vert \left\{  a\right\}  \cap\left\{  a,c\right\}  \right\vert
}{\left\vert \left\{  a,c\right\}  \right\vert }=\frac{1}{2}$;\\
$\Pr(\left\{  b\right\}  $ measured at wall $|\left\{  a,c\right\}  $ at
slits$)=\Pr\left(  \left\{  b\right\}  |\left\{  a,c\right\}  \right)
=\frac{\left\vert \left\{  b\right\}  \cap\left\{  a,c\right\}  \right\vert
}{\left\vert \left\{  a,c\right\}  \right\vert }=0$;\\
$\Pr(\left\{  c\right\}  $ measured at wall $|\left\{  a,c\right\}  $ at
slits$)=\Pr\left(  \left\{  c\right\}  |\left\{  a,c\right\}  \right)
=\frac{\left\vert \left\{  c\right\}  \cap\left\{  a,c\right\}  \right\vert
}{\left\vert \left\{  a,c\right\}  \right\vert }=\frac{1}{2}$.
\end{tabular}
%

%TCIMACRO{\FRAME{dtbpF}{2.5139in}{1.8032in}{0in}{}{}{fig7-case2-two-slit.eps}%
%{\special{ language "Scientific Word";  type "GRAPHIC";
%maintain-aspect-ratio TRUE;  display "USEDEF";  valid_file "F";
%width 2.5139in;  height 1.8032in;  depth 0in;  original-width 2.3993in;
%original-height 1.7136in;  cropleft "0";  croptop "1";  cropright "1";
%cropbottom "0";  filename 'fig7-case2-two-slit.eps';file-properties "XNPEU";}}
%}%
%BeginExpansion
\begin{center}
\includegraphics[
height=1.8032in,
width=2.5139in
]%
{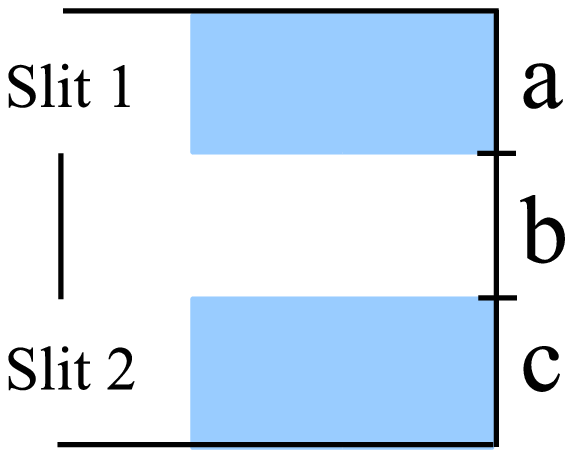}%
\end{center}
%EndExpansion

Figure 7: Final probability distribution with no measurement at slits
\end{center}

Since no "collapse" took place at the slits due to no distinctions being made
there, the indistinct element $\left\{  a,c\right\}  $ evolved (rather than
one or the other of the distinct elements $\left\{  a\right\}  $ or $\left\{
c\right\}  $). The action of $A$ is the same on $\left\{  a\right\}  $ and
$\left\{  c\right\}  $ as when they evolve separately since $A$ is a linear
operator but the two results are now added together \textit{as part of the
evolution}. This allows the "interference" of the two results and thus the
cancellation of the $\left\{  b\right\}  $ term in $\left\{  a,b\right\}
+\left\{  b,c\right\}  =\left\{  a,c\right\}  $. The addition is, of course,
mod $2$ (where $-1=+1$) so, in "wave language," the two "wave crests" that add
at the location $\left\{  b\right\}  $ cancel out. When this indistinct
element $\left\{  a,c\right\}  $ "hits the wall" on the right, there is an
equal probability of that distinction yielding either of those eigenstates.
Figure 7 shows the simplest example of the "light and dark bands"
characteristic of superposition and interference illustrating "all of the
mystery of quantum mechanics".

This pedagogical model gives the simple logical essence of the two-slit
experiment without the complex-valued wave functions that distract from the
essential point--which is the difference between the separate mixed state
evolutions resulting from measurement at the slits, and the combined evolution
of the superposition $\left\{  a,c\right\}  $ that allows interference
(without "waves").

\section{Further steps}

Showing that ordinary Laplace-Boole finite probability theory is the quantum
probability calculus for the pedagogical or "toy" model, quantum mechanics
over sets (QM/sets), is only an initial part of a research programme. For
instance, we have not considered:

\begin{itemize}
\item the whole "non-commutative" side of viewing the Laplace-Boole theory in
the context of vector spaces over $%
%TCIMACRO{\U{2124} }%
%BeginExpansion
\mathbb{Z}
%EndExpansion
_{2}$ where the compatibility of numerical attributes $f:U\rightarrow%
%TCIMACRO{\U{211d} }%
%BeginExpansion
\mathbb{R}
%EndExpansion
$ and $g:U^{\prime}\rightarrow%
%TCIMACRO{\U{211d} }%
%BeginExpansion
\mathbb{R}
%EndExpansion
$ defined on different equicardinal basis sets $\left\{  u\right\}  \subseteq
U$ and $\left\{  u^{\prime}\right\}  \subseteq U^{\prime}$ can be analyzed in
terms of the commutativity of all the associated projection operators
$f^{-1}\left(  r\right)  \cap()$ and $g^{-1}\left(  s\right)  \cap()$ on $%
%TCIMACRO{\U{2124} }%
%BeginExpansion
\mathbb{Z}
%EndExpansion
_{2}^{n}$ \cite{ell:objindef}; or

\item the treatment of entanglement in QM/sets which reduces to some
old-fashioned correlation in the equiprobability distribution on a state that
is a subset of a Cartesian product but which still allows a Bell-type result
to be established \cite{ell:qmoversets}.
\end{itemize}

Our purpose here is limited to showing how the perfectly classical
Laplace-Boole finite probability theory is the quantum probability calculus of
the pedagogical model of quantum mechanics over sets. The point is not to
clarify finite probability theory but to elucidate quantum mechanics itself by
seeing some of its quantum features formulated in a classical setting.

\end{document}